\documentclass{article}

\usepackage{arxiv}

\usepackage[utf8]{inputenc} % allow utf-8 input
\usepackage[T1]{fontenc}    % use 8-bit T1 fonts
\usepackage{hyperref}       % hyperlinks
\usepackage{url}            % simple URL typesetting
\usepackage{booktabs}       % professional-quality tables
\usepackage{amsfonts}       % blackboard math symbols
\usepackage{nicefrac}       % compact symbols for 1/2, etc.
\usepackage{microtype}      % microtypography
\usepackage{lipsum}		% Can be removed after putting your text content
\usepackage{graphicx}
\usepackage{natbib}
\usepackage{doi}

\usepackage{amsmath}	% Advanced maths commands
\usepackage{multirow}
\usepackage{graphicx}
\usepackage{booktabs}
\usepackage{longtable}
\usepackage{supertabular}
\usepackage{caption}
\usepackage{lineno}

\newcommand{\aapr}{AApR}
\newcommand{\apj}{ApJ}

\title{Theoretical Study of Inelastic Processes in Collisions of Y and Y$^+$ with Hydrogen Atom}

\author{ 
        \href{https://orcid.org/0000-0002-2448-3049}{\includegraphics[scale=0.06]{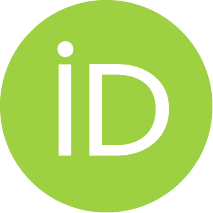}\hspace{1mm}Yu Wang}\\
	School of Physics,\\
        Beijing Institute of Technology\\
	Beijing, 100081, China \\
	\texttt{wangyu\_1899@126.com} \\
	\And
	\href{https://orcid.org/0000-0002-8709-4665}{\includegraphics[scale=0.06]{orcid.pdf}\hspace{1mm}Sofya Alexeeva} \\
	CAS Key Laboratory of Optical Astronomy\\
	National Astronomical Observatories\\
	 Beijing, 100101, China \\
        \And
	\href{https://orcid.org/0000-0002-8514-4497}{\includegraphics[scale=0.06]{orcid.pdf}\hspace{1mm}Feng Wang} \\
	 School of Physics,\\
        Beijing Institute of Technology\\
	Beijing, 100081, China \\
	 \And
	 \href{https://orcid.org/0000-0001-7816-4804}{\includegraphics[scale=0.06]{orcid.pdf}\hspace{1mm}Ling Liu} \\
	 Data Center for High Energy Density Physics\\
	 Institute of Applied Physics and Computational Mathematics\\
	 Beijing 100088, China\\
	 \texttt{liu\_ling@iapcm.ac.cn} \\
	 \And
	  \href{https://orcid.org/0000-0003-1874-9653}{\includegraphics[scale=0.06]{orcid.pdf}\hspace{1mm}Yong Wu} \\
	  Data Center for High Energy Density Physics\\
	 Institute of Applied Physics and Computational Mathematics\\
	 Beijing 100088, China\\
	  \texttt{wu\_yong@iapcm.ac.cn} \\
	 \And
	 {JianGuo Wang} \\
	 Data Center for High Energy Density Physics\\
	 Institute of Applied Physics and Computational Mathematics\\
	 Beijing 100088, China\\
	 \And
	 {Gang Zhao} \\
	 CAS Key Laboratory of Optical Astronomy\\
	National Astronomical Observatories\\
	 Beijing, 100101, China \\
	 \And
	 \href{https://orcid.org/0000-0002-8889-7283}{\includegraphics[scale=0.06]{orcid.pdf}\hspace{1mm} Svetlana A. Yakovleva} \\
	 Department of Theoretical Physics and Astronomy\\
	 Herzen University\\
	 St Petersburg 191186, Russia\\
	  \And
	  \href{https://orcid.org/0000-0001-8834-1456}{\includegraphics[scale=0.06]{orcid.pdf}\hspace{1mm}Andrey K. Belyaev} \\
	 Department of Theoretical Physics and Astronomy\\
	 Herzen University\\
	 St Petersburg 191186, Russia\\
}
\date{\today}

\renewcommand{\shorttitle}{\textit{arXiv} Template}

\hypersetup{
pdftitle={A template for the arxiv style},
pdfsubject={q-bio.NC, q-bio.QM},
pdfauthor={David S.~Hippocampus, Elias D.~Striatum},
pdfkeywords={First keyword, Second keyword, More},
}

\begin{document}
\maketitle

\begin{abstract}
	Utilizing a simplified quantum model approach, the low-energy inelastic collision processes between yttrium atoms (ions) and hydrogen atoms have been studied. Rate coefficients corresponding to the mutual neutralization, ion-pair formation, excitation, and de-excitation processes for the above collision systems have been provided in the temperature range of 1000-10000K. 3 ionic states and 73 covalent states are considered in calculations for the collisions of yttrium atoms with hydrogen atoms, which include 6 molecular symmetries and 4074 partial inelastic reaction processes. For the collisions of yttrium ions with hydrogen atoms, 1 ionic state and 116 covalent states are included, which related to 3 molecular symmetries and 13572 partial inelastic collision processes. It is found that the rate coefficients for the mutual neutralization process have a maximum at T = 6000K, which is an order of magnitude higher than those of other processes. Notably, the positions of optimal windows for the collisions of yttrium atoms and ions with hydrogen atoms are found near electronic binding energy -2eV (Y) and -4.4eV (Y$^+$), respectively. The scattering channels located in or near these optimal windows have intermediate-to-large rate coefficients (greater than $10^{-12}$ cm$^3$s$^{-1}$). The reported data should be useful in the study of non-local thermodynamic equilibrium modeling. 
\end{abstract}

% keywords can be removed

\section{Introduction}
The abundance and distribution of the Yttrium element significantly influence the investigations of the chemical evolution of the cosmos, particularly within stellar environments. This light neutron-capture element is readily observed in stars of B to K-type, and it is primarily produced via three recognized neutron-capture reactions: the rapid ($r$-) process, the main component of the slow ($s$-) process, and the weak component of the $s$-process \citep{Kappeler1989}. These processes occur in stars with different types and masses \citep{Gallino1998, Kappeler2011}, offering a unique perspective to understand the enrichment history of heavy elements within the interstellar medium, and providing opportunities to refine the theories of nucleosynthesis.   

The investigation of Yttrium abundance is highly relevant to stars with various spectral types and populations. For thin disk stars and solar-type stars, an accurate estimation of Yttrium abundance, in conjunction with other spectroscopic indices, can provide a tool to estimate the ages of stars. This has always been a challenge since the inception of stellar astrophysics \citep{Silva2012, Nissen2015, Tucci2016, Titarenko2019, Berger2022}. Moreover, the abundance ratio of Yttrium to Europium ([Y/Eu]) can serve as an efficient indicator of the efficiency of chemical evolution; it can also characterize the relative contribution of low- to intermediate-mass stars compared to high-mass stars \citep{Recio2021}.

In order to better understand the origin of elements, the Galactic chemical evolution (GCE) models for these elements are constructed (see e.g. \citet{2020Kobayashi}). The observed abundance trends of [X/Fe] versus [Fe/H] for particular chemical element X provide observational constraints GCE models. For yttrium element, numerous studies has been dedicated to establishing [Y/Fe] vs. [Fe/H] Galactic trend 
(e.g. \citet{Johnson2002, honda2004, francois2007, hansen2012, roederer2014, bensby2014, reggiani2017, baratella2021, li2022}).
These studies have revealed a moderate underabundance of the yttrium element in stars where the value of [Fe/H] is less than -1, accompanied by significant dispersions in the [Y/Fe] ratio.  However, the yttrium abundances in these studies were obtained under the assumption of local thermodynamic equilibrium (LTE), which may lead to inaccurate results, especially, in metal-poor stellar atmospheres.

The non-local thermodynamic equilibrium (NLTE) method is a realistic alternative \citep{Asplund2005, Barklem2016}. This method can reveal important information about stellar composition and provide abundance data for many elements. A large amount of precise atomic data related to the element is required for the modeling in the NLTE method, including atomic energy levels, radiation data, collision data with electrons, and collision data with hydrogen atoms. The greatest uncertainty in NLTE modeling arises from the atomic data of non-elastic processes in ion/atom-hydrogen collisions \citep{Asplund2005, Mashonkina2014, Barklem2016}. In previous researches, the quantum cross-sections for non-elastic collisions between different atoms and hydrogen atoms have traditionally been estimated using the Drawin formula, which is derived from the classical Thomson model. Nevertheless, studies by \citet{Barklem2011} indicated that the Drawin formula lacked an accurate physical foundation for low-energy collisions and cannot provide reliable results. Therefore, it is necessary to replace the Drawin formula with a reasonable quantum calculation to obtain the accurate atomic data for collisions involving hydrogen atoms.

Recently, the full quantum mechanical method has been employed to study the collisions between complex neutral atoms (or ions) and hydrogen atoms at low collision energies. Due to the complexity of the collision systems, these computations are exceptionally time-consuming. This approach is only suitable for dealing with simple collision systems (including a few electrons) or complex collision systems that have been simplified by freezing their inner layer electrons, such as Li+H (Li$^+$+H$^-$)\citep{Croft1999a, Croft1999b, Belyaev2003}, Na+H (Na$^+$+H$^-$) \citep{Belyaev1999, Belyaev2010}, Mg+H (Mg$^+$+H$^-$) \citep{Belyaev2012, Guitou2015}, Ca+H (Ca$^+$+H$^-$) \citep{Mitrush2017, Belyaev2019}. However, for low-energy non-elastic collisions between hydrogen atoms and other complex atoms (or ions), the non-adiabatic couplings only contribute to the cross sections at large inter-nuclear distances. Based on this observation, Belyaev et al. proposed several quantum approximation models, including the quantum multi-channel analysis method \citep{Belyaev1993}, the quantum branching probability current method \citep{Belyaev2013a}, and a simplified quantum model \citep{Belyaev2017a, Belyaev2017b} to deal with the complex collision systems. Comparing with the full quantum mechanical calculations, these models are reasonable in physics and the computational difficulty is greatly reduced, making them appropriate for estimating the cross-sections and rate coefficients for non-elastic collisions between hydrogen atoms and other complex atoms (or ions) at low collision energies.

In these quantum approximation models, the simplified quantum model employs basic Coulomb and flat potentials to represent the ionic and covalent states of the collision sysetms, respectively. The transition probabilities between different reaction channels are described by directly related non-adiabatic probabilities. This streamlined structure and non-adiabatic dynamics method make it more convenient and efficient to deal with the complex heavy particle collision systems compared with the other two quantum approximation models \citep{Belyaev1993, Belyaev2013a}. Moreover, the simplified model can also provide reliable rate coefficients of intermediate-to-large magnitude and reasonable estimations for the rate coefficients with small values comparing with the full quantum calculations. This model has been successfully implemented in several complex collision systems, such as Fe+H (Fe$^+$+H$^-$) \citep{Yakovleva2018, Yakovleva2019}, Co+H (Co$^+$+H$^-$) \citep{Yakovleva2020} and Ni+H \citep{Voronov2022}. The simplified model can serves as an effective tool for studying the non-elastic collision processes between complex atoms and hydrogen atoms in the cases which relevant quantum chemistry data is lacking. The motivation of the present work is to provide the rate coefficients for the non-elastic collision processes between yttrium atoms (and ions) and hydrogen atoms at low energies by employing the simplified model based on the Landau-Zener method \citep{Belyaev2017a, Belyaev2017b}.

The organization of this paper is as follows: Section \ref{S2} outlines the details of the simplified model that we use. Section \ref{S3} discusses the results of our rate-coefficient calculations for the non-elastic collision processes in yttrium (and yttrium ions)-hydrogen collisions. Our conclusions are summarized in Section \ref{S4}.

\section{Simplified model} \label{S2}
The quantum studies of the collisions between the neutral atoms and hydrogen atoms indicate that the interaction between ionic and covalent molecular states provides the primary mechanism for non-adiabatic transitions. This long-range ionic-covalent interaction mechanism plays a critical role in non-elastic collisions (such as charge transfer, mutual neutralization, ion-pair formation, and excitation and de-excitation processes.) involving hydrogen atoms. Due to the importance of this mechanism, \citet{Belyaev2017a, Belyaev2017b} have proposed a simplified model to estimate relevant rate coefficients for the above collision systems. Here, we provide a brief overview of this model.

The simplified model which based on a semi-empirical ionic-covalent interaction theory, determines the long-range electronic structure of the collision systems. In the A$^{Z+}(j)$+H and A$^{(Z+1)+}$+H$^-$ collision systems, where A represents the considered chemical element, $Z$=0 represents  collisions of neutral atoms with hydrogen atoms, and $Z$$\le$1  denotes the collisions between ions and hydrogen atoms. In order to estimate the state-to-state transition probabilities which leading to the large rate coefficients, we need to know how to represent the long-range adiabatic molecular potentials of the ionic state A$^{(Z+1)+}$+H$^-$ and the covalent state A$^{Z+}(j)$+H, by $H_{\mathrm{ionic,ionic}}$ and $H_{jj}$, respectively, varying with the nuclear distance $R$. The molecular potential of the ionic state can be expressed by the Coulomb potential as, 
\begin{equation}
H_{\mathrm{ionic,ionic}}(R)=E_{\mathrm{H}^{-}}-\frac{Z+1}{R}.
\end{equation}
where $E_{H^-}$=-0.754eV, is electronic bound energy of hydrogen anion. The potential energy of the covalent molecular state can be described by a flat potential as \begin{equation}
H_{jj}(R)=E_{j}.
\end{equation}

where electronic bound energy $E_j$ is defined as $\widetilde{E}_j$ minus $I_{\text{ionizaiton}}$, where $I_{\text{ionizaiton}}$ is the ionization energy of A$^{Z+}$, $\widetilde{E}_j$ is the electronic energy taken from the ground state of A$^{Z+}$ (considered zero energy). The off-diagonal matrix element can be estimated using the potential energies of the covalent and ionic states. For the single-electron capture process, $H_{\mathrm{ionic} j}$ can be determined using the semi-empirical formula proposed by \citet{Olson1971}.
\begin{equation}
\begin{aligned}
H_{\text {ionic } j}= & \frac{\sqrt{-E_j}+\sqrt{-E_{\mathrm{H}^{-}}}}{\sqrt{2}} \sqrt{E_j E_{\mathrm{H}^{-}}} R_j \\
& \times \exp  \left(-0.86 R_j \frac{\sqrt{-E_j}+\sqrt{-E_{\mathrm{H}^{-}}}}{\sqrt{2}} \right) .
\end{aligned}
\end{equation}

 where $R_j$ is the center of the nonadiabatic region. The non-adiabatic dynamics of the nucleus are studied using the Landau-Zener model, and the non-adiabatic transition probabilities $p_j$ can be calculated based on the diabatic potential energy curves as 
\begin{equation}
p_j=p_{\mathrm{ionic} j}=\exp  \left(-\frac{2 \pi H_{\text {ionic } j}^2(Z+1)}{ (E_{\mathrm{H}^{-}}-E_j )^2 \hbar v} \right).
\label{e4}
\end{equation}
The cross section can be obtained by summing the total angular momentum quantum number $J$.
\begin{equation}
\sigma_{if}^{\Lambda}(E)=\frac{\pi \hbar^2p_i^{\text {stat}}}{2 \mu E} \sum_{J=0}^{J_{\max }} P_{i f}^{\Lambda}(J, E)(2 J+1).
\end{equation}
Herein, $p_i^{\text {stat}}$ denotes the statistical probability of the initial molecular state $i$, and $\Lambda$ represents the different molecular symmetries. $P_{if}$ represents the transition probability from state $i$ to state $f$, which has different expressions for mutual neutralization and de-excitation processes, respectively. The simplified method ignores the effects of other non-adiabatic regions, and the transition probability can only be represented by the directly related non-adiabatic probability $p_k$. In the mutual neutralization process, an incident projectile from the initial ionic state passes through the non-adiabatic region $R_f$, and is then emitted through a covalent state channel $f$. The transition probability can be expressed as $P_{i f}^{N,{\Lambda}}=2p_f(1-p_f)$. Meanwhile, in the de-excitation process, an incident projectile from the initial covalent channel $i$ passes through the non-adiabatic regions $R_i$ and $R_f$, and is then emitted from the covalent channel $f$, with the transition probability presented as $P_{i f}^{D,{\Lambda}}=2p_f(1-p_f)(1-p_i)$. It's important to note that the expression of the velocity $v$ in Equation \eqref{e4} is different for these two collision processes \citet{Belyaev2017b}.
\begin{equation}
\begin{aligned}
& v=v_f^N=\sqrt{\frac{2}{\mu}\left(E+E_{\mathrm{H}^{-}}-E_f-\frac{J(J+1) \hbar^2}{2 \mu R_f^2}\right)} . \\
& v=v_i^D=\sqrt{\frac{2}{\mu}\left(E-\frac{J(J+1) \hbar^2}{2 \mu R_i^2}\right)}. \\
& v=v_f^D=\sqrt{\frac{2}{\mu}\left(E+E_i-E_f-\frac{J(J+1) \hbar^2}{2 \mu R_f^2}\right)} .
\end{aligned}
\end{equation}
Finally, the rate coefficients for both neutralization and de-excitation processes at various temperatures can be determined through the Maxwell-Boltzmann integration over the partial cross section as follows,
\begin{equation}
K_{if}^{\Lambda}(T)=\sqrt{\frac{8}{\pi \mu (k_{\mathrm{B}} T )^3}} \int_0^{\infty} E \sigma_{i f}^{\Lambda}(E) \exp \left(-\frac{E}{k_{\mathrm{B}} T} \right) \mathrm{d} E.
\end{equation}
where $K_B$ is Boltzmann constant. Neglecting the influence of $p_i^{\text {stat}}$, it can be observed that the rate coefficient for the transition process $i$$\rightarrow$$f$ is only related to a covalent state binding energy $E_f$ in the neutralization process.  For de-excitation process, the rate coefficient is only related to the binding energies ($E_i$ and $E_f$) of two covalent states. So the reduced rate coefficient for the mutual neutralization and de-excitation processes can be expressed as a function of electron binding energy as,
\begin{equation}
\begin{aligned}
N_{i f}^{\Lambda}(T ; E_f)&=\sqrt{\frac{8}{\pi \mu (k_{\mathrm{B}} T )^3}} \int_0^{\infty} \sigma_{i f}^{N, \Lambda}(E) E \exp  \left(-\frac{E}{k_{\mathrm{B}} T} \right) \mathrm{d} E.\\
D_{i f}^{\Lambda}(T ; E_i, E_f)&=\sqrt{\frac{8}{\pi \mu (k_{\mathrm{B}} T )^3}} \int_0^{\infty} \sigma_{i f}^{D, \Lambda}(E) E \exp  \left(-\frac{E}{k_{\mathrm{B}} T} \right) \mathrm{d} E.
\end{aligned}
\end{equation}
$N_{if}$ and $D_{if}$ are defined as the reduced rate coefficients of the mutual neutralization and de-excitation processes, respectively.
Then the total rate coefficient for state-to-state transitions can be obtained by considering various molecular symmetries and spins $S$ as
\begin{equation}
\begin{aligned}
K_{if}(T)&=\sum_{2 S+1,\Lambda} p_i^{\text {stat},2S+1,\Lambda} N_{if}^{2S+1,\Lambda}(T ; E_f).\\
K_{i f}(T)&=\sum_{2 S+1,\Lambda} p_i^{\text {stat },2S+1,\Lambda} D_{i f}^{2 S+1,\Lambda}(T ; E_i, E_f).
\end{aligned}
\end{equation}
Employing the detailed balance relation, the rate coefficients for ion-pair formation and excitation processes can be deduced from the mutual neutralization and de-excitation processes, respectively, as 
\begin{equation}
K_{f i}(T)=K_{i f}(T) \frac{p_f^{\text {stat }}}{p_i^{\text {stat }}} \exp  \left(-\frac{\Delta E_{i f}}{k_{\mathrm{B}} T} \right).
\end{equation}
where $\Delta E_{if}= E_i-E_j$, is greater than 0.

\section{INELASTIC COLLISIONS WITH HYDROGEN}  \label{S3}
\subsection{Y+H and Y$^+$+H$^-$ collisions}

\begin{table}
 \setlength{\tabcolsep}{2pt}
 \renewcommand{\arraystretch}{1} % 
\centering   
\caption{
YH Molecular States for Yttrium Ionic Configuration Y$^+(5s^2 \mathrm{a}^1\mathrm{S})$ and Y$^+(4d5s \mathrm{a}^3\mathrm{D})$, the Corresponding Asymptotic Atomic States (Scattering Channels), the Asymptotic Energies ($J$- averaged values taken from NIST \citep{Palmer1977}) and the Considered Molecular Symmetries for the Treated Molecular States. 
}             
\label{table1}             
\begin{tabular}{ l l l l l l l }    % 7 columns

\hline \hline
\multirow{1}{*} {j} &\multirow{1}{*}  {Asymptotic atomic states} & \multirow{1}{*} {Asymptotic }& \multicolumn{4}{c} { Molecular  } \\
\multirow{1}{*}     &\multirow{1}{*}                                         & \multirow{1}{*} {energies (eV) }& \multicolumn{4}{c} {  symmetry} \\
 \hline
 1 & $\mathrm{Y}(4d5s^2$ $\mathrm{a}^2\mathrm{D} )+\mathrm{H} (1s$ $^2\mathrm{S})$                                                  & 0.00000000 & $^1\Sigma^+$  & $^3\Sigma^+$  & $^3\Pi$  & $^3\Delta$  \\
 2 & $\mathrm{Y}(5s^25p$ $\mathrm{z}^2\mathrm{P}^{\mathrm{o}} )+\mathrm{H} (1s$ $^2\mathrm{S})$                              & 1.30545058 & $^1\Sigma^+$  & $^3\Sigma^+$  & $^3\Pi$  & \\
 3 & $\mathrm{Y}(4d5s(^3\mathrm{D})5p$ $\mathrm{z}^4\mathrm{F}^{\mathrm{o}} )+\mathrm{H} (1s$ $^2\mathrm{S})$        & 1.85343904 & $^1\Sigma^+$  & $^3\Sigma^+$  & $^3\Pi$  & $^3\Delta$  \\
 4 & $\mathrm{Y}(4d^2(^1\mathrm{D})5s$ $\mathrm{b}^2\mathrm{D})+\mathrm{H} (1s$ $^2\mathrm{S})$                            & 1.98300885  & $^1\Sigma^+$  & $^3\Sigma^+$  & $^3\Pi$  & $^3\Delta$  \\
 5 & $\mathrm{Y}(4d^2(^1\mathrm{G})5s$ $\mathrm{a}^2\mathrm{G})+\mathrm{H} (1s$ $^2\mathrm{S})$                            & 2.29362274  & $^1\Sigma^+$  & $^3\Sigma^+$  & $^3\Pi$  & $^3\Delta$  \\
 6 & $\mathrm{Y}(4d5s(^3\mathrm{D})5p$ $\mathrm{z}^4\mathrm{P}^{\mathrm{o}})+\mathrm{H} (1s$ $^2\mathrm{S})$        & 2.35276953 & $^1\Sigma^+$  & $^3\Sigma^+$  & $^3\Pi$  & \\
 7 & $\mathrm{Y}(4d5s(^3\mathrm{D})5p$ $\mathrm{z}^2\mathrm{F}^{\mathrm{o}} )+\mathrm{H} (1s$ $^2\mathrm{S})$       & 2.66920361 & $^1\Sigma^+$  & $^3\Sigma^+$  & $^3\Pi$  & $^3\Delta$  \\
 8 & $\mathrm{Y}(4d^2(^1\mathrm{S})5s$ $^2\mathrm{S})+\mathrm{H} (1s$ $^2\mathrm{S})$                                            & 2.90930070 & $^1\Sigma^+$  & $^3\Sigma^+$  & & \\
 9 & $\mathrm{Y}(4d5s(^3\mathrm{D})5p$ $\mathrm{y}^2\mathrm{P}^{\mathrm{o}} )+\mathrm{H} (1s$ $^2\mathrm{S})$       & 3.03521575 & $^1\Sigma^+$  & $^3\Sigma^+$  & $^3\Pi$  & \\
 10 & $\mathrm{Y}(4d5s(^1\mathrm{D})5p$ $\mathrm{y}^2\mathrm{F}^{\mathrm{o}})+\mathrm{H} (1s$ $^2\mathrm{S})$       & 3.03993769 & $^1\Sigma^+$  & $^3\Sigma^+$  & $^3\Pi$  & $^3\Delta$  \\
 11 & $\mathrm{Y}(4d5s(^1\mathrm{D})5p$ $\mathrm{x}^2\mathrm{P}^{\mathrm{o}})+\mathrm{H} (1s$ $^2\mathrm{S})$       & 3.44981660 & $^1\Sigma^+$  & $^3\Sigma^+$  & $^3\Pi$  & \\
 12 & $\mathrm{Y}(4d^2(^3\mathrm{F})5p$ $\mathrm{y}^4\mathrm{F}^{\mathrm{o}})+\mathrm{H} (1s$ $^2\mathrm{S})$      & 3.90654880 & $^1\Sigma^+$  & $^3\Sigma^+$  & $^3\Pi$  & $^3\Delta$  \\
 13 & $\mathrm{Y}(5s^26s$ $\mathrm{e}^2\mathrm{S} )+\mathrm{H} (1s$ $^2\mathrm{S})$                                             & 3.92677040 & $^1\Sigma^+$  & $^3\Sigma^+$  & & \\
 14 & $\mathrm{Y}({4d^3}_1$ $^2\mathrm{D})+\mathrm{H} (1s$ $^2\mathrm{S})$                                                            & 4.02102640 & $^1\Sigma^+$  & $^3\Sigma^+$  & $^3\Pi$  & $^3\Delta$  \\
 15 & $\mathrm{Y}(4d5s(^3\mathrm{D})6s$ $\mathrm{e}^4\mathrm{D})+\mathrm{H} (1s$ $^2\mathrm{S})$                        & 4.10988177 & $^1\Sigma^+$  & $^3\Sigma^+$  & $^3\Pi$  & $^3\Delta$  \\
 16 & $\mathrm{Y}(4d^2(^3\mathrm{F})5p$ $\mathrm{x}^2\mathrm{F}^{\mathrm{o}})+\mathrm{H} (1s$ $^2\mathrm{S})$    & 4.16689962 & $^1\Sigma^+$  & $^3\Sigma^+$  & $^3\Pi$  & $^3\Delta$  \\
 17 & $\mathrm{Y}(5s^2(^2\mathrm{D})5d$ $\mathrm{e}^2\mathrm{D})+\mathrm{H} (1s$ $^2\mathrm{S})$                       & 4.24417054 & $^1\Sigma^+$  & $^3\Sigma^+$  & $^3\Pi$  & $^3\Delta$  \\
 18 & $\mathrm{Y}(4d5s(^3\mathrm{D})6s$ $\mathrm{f}^2\mathrm{D})+\mathrm{H} (1s$ $^2\mathrm{S})$                       & 4.51558026 & $^1\Sigma^+$  & $^3\Sigma^+$  & $^3\Pi$  & $^3\Delta$  \\
 19 & $\mathrm{Y}(4d5s(^1\mathrm{D})6s$ $\mathrm{f}^2\mathrm{D})+\mathrm{H} (1s$ $^2\mathrm{S})$                        & 4.57010530 & $^1\Sigma^+$  & $^3\Sigma^+$  & $^3\Pi$  & $^3\Delta$  \\
 20 & $\mathrm{Y}(4d^2(^3\mathrm{P})5p$ $\mathrm{y}^4\mathrm{P}^{\mathrm{o}})+\mathrm{H} (1s$ $^2\mathrm{S})$   & 4.59228610 & $^1\Sigma^+$  & $^3\Sigma^+$  & $^3\Pi$  & \\
 21 & $\mathrm{Y}(5s^2(^1\mathrm{S})6p$ $^2\mathrm{P}^{\mathrm{o}})+\mathrm{H} (1s$ $^2\mathrm{S})$                    & 4.60052330 & $^1\Sigma^+$  & $^3\Sigma^+$  & $^3\Pi$  & \\
 22 & $\mathrm{Y}(4d^2(^1\mathrm{D})5p$ $\mathrm{w}^2\mathrm{P}^{\mathrm{o}})+\mathrm{H} (1s$ $^2\mathrm{S})$   & 4.61764050 & $^1\Sigma^+$  & $^3\Sigma^+$  & $^3\Pi$  & \\
 23 & $\mathrm{Y}(4d^2(^1\mathrm{D})5p$ $\mathrm{w}^2\mathrm{F}^{\mathrm{o}})+\mathrm{H} (1s$ $^2\mathrm{S})$   & 4.62202960 & $^1\Sigma^+$  & $^3\Sigma^+$  & $^3\Pi$  & $^3\Delta$  \\
 24 & $\mathrm{Y}({4d^3}_2$ $^2\mathrm{D})+\mathrm{H} (1s$ $^2\mathrm{S})$                                                         & 4.65243770 & $^1\Sigma^+$  & $^3\Sigma^+$  & $^3\Pi$  & $^3\Delta$  \\
 25 & $\mathrm{Y}(4d^2(^1\mathrm{G})5p$ $\mathrm{z}^2\mathrm{H}^{\mathrm{o}})+\mathrm{H} (1s$ $^2\mathrm{S})$   & 4.66033640 & $^1\Sigma^+$  & $^3\Sigma^+$  & $^3\Pi$  & $^3\Delta$  \\
 26 & $\mathrm{Y}(4d5s(^3\mathrm{D})5d$ $\mathrm{f}^4\mathrm{D})+\mathrm{H} (1s$ $^2\mathrm{S})$                      & 4.76966640 & $^1\Sigma^+$  & $^3\Sigma^+$  & $^3\Pi$  & $^3\Delta$  \\
 27 & $\mathrm{Y}(4d5s(^3\mathrm{D})5d$ $\mathrm{e}^4\mathrm{G})+\mathrm{H} (1s$ $^2\mathrm{S})$                     & 4.79018910 & $^1\Sigma^+$  & $^3\Sigma^+$  & $^3\Pi$  & $^3\Delta$  \\
 28 & $\mathrm{Y}(4d5s(^3\mathrm{D})6p$ $^4\mathrm{F}^{\mathrm{o}})+\mathrm{H} (1s$ $^2\mathrm{S})$                 & 4.79813880 & $^1\Sigma^+$  & $^3\Sigma^+$  & $^3\Pi$  & $^3\Delta$  \\
 29 & $\mathrm{Y}(4d5s(^3\mathrm{D})6p$ $^2\mathrm{F}^{\mathrm{o}})+\mathrm{H} (1s$ $^2\mathrm{S})$                 & 4.82157090 & $^1\Sigma^+$  & $^3\Sigma^+$  & $^3\Pi$  & $^3\Delta$  \\
 30 & $\mathrm{Y}(4d5s(^3\mathrm{D})5d$ $^4\mathrm{S} )+\mathrm{H} (1s$ $^2\mathrm{S})$                                   & 4.83556540 & $^1\Sigma^+$  & $^3\Sigma^+$  &  &  \\
 31 & $\mathrm{Y}(4d5s(^3\mathrm{D})6p$ $^4\mathrm{P}^{\mathrm{o}})+\mathrm{H} (1s$ $^2\mathrm{S})$                 & 4.86903590 & $^1\Sigma^+$  & $^3\Sigma^+$  & $^3\Pi$  & \\
 32 & $\mathrm{Y}(4d5s(^3\mathrm{D})5d$ $^2\mathrm{D})+\mathrm{H} (1s$ $^2\mathrm{S})$                                    & 5.13578369 & $^1\Sigma^+$  & $^3\Sigma^+$  & $^3\Pi$  & $^3\Delta$  \\
 33 & $\mathrm{Y}(4d5s(^1\mathrm{D})5d$ $^2\mathrm{D})+\mathrm{H} (1s$ $^2\mathrm{S})$                                    & 5.16524630& $^1\Sigma^+$  & $^3\Sigma^+$  & $^3\Pi$  & $^3\Delta$  \\
 34 & $\mathrm{Y}(5p^2(^1\mathrm{D})5s$ $^2\mathrm{D})+\mathrm{H} (1s$ $^2\mathrm{S})$                                     & 5.21340772 & $^1\Sigma^+$  & $^3\Sigma^+$  & $^3\Pi$  & $^3\Delta$  \\
 35 & $\mathrm{Y}(4d^2(^1\mathrm{G})5p$ $\mathrm{v}^2\mathrm{F}^{\mathrm{o}} )+\mathrm{H} (1s$ $^2\mathrm{S})$ & 5.31369680 & $^1\Sigma^+$  & $^3\Sigma^+$  & $^3\Pi$  & $^3\Delta$  \\
 36 & $\mathrm{Y}(5s^27s$ $\mathrm{g}^2\mathrm{S})+\mathrm{H} (1s$ $^2\mathrm{S})$                                           & 5.41111690 & $^1\Sigma^+$  & $^3\Sigma^+$  &  & \\
 \hline
 37 & $\mathrm{Y}^{+}(5s^2$ $\mathrm{a}^1\mathrm{S})+\mathrm{H}^{-} (1s^2$ $^1\mathrm{ S})$                                          & 5.46326000 & $^1\Sigma^+$ & & & \\
 \hline
 38 & $\mathrm{Y}(4d 5s(^3\mathrm{D})7s$ $\mathrm{g}^4\mathrm{D})+\mathrm{H} (1s$ $^2\mathrm{S})$                    & 5.53649000 & $^1\Sigma^+$  & $^3\Sigma^+$  & $^3\Pi$  & $^3\Delta$  \\
 39 & $\mathrm{Y}(5p^2(^3\mathrm{P})4d$ $\mathrm{ h}^4\mathrm{D} )+\mathrm{H} (1s$ $^2\mathrm{S})$                  & 5.53713880 & $^1\Sigma^+$  & $^3\Sigma^+$  & $^3\Pi$  & $^3\Delta$  \\
 \hline
 40 & $\mathrm{Y}^{+} (4d5s$ $\mathrm{a}^3\mathrm{D})+\mathrm{H}^{-} (1s^2$ $^1\mathrm{ S})$                              & 5.56743130 &  & $^3\Sigma^+$  & $^3\Pi$  & $^3\Delta$  \\
\hline
\end{tabular}
\end{table}

\begin{table} 
 \setlength{\tabcolsep}{2pt}
 \renewcommand{\arraystretch}{1} % 
\centering   
\caption{
YH Molecular States for Yttrium Ionic Configuration Y$^+(5s^2 \mathrm{a}^1\mathrm{S})$ and Y$^+(4d5s \mathrm{a}^3\mathrm{D})$, the Corresponding Asymptotic Atomic States (Scattering Channels), the Asymptotic Energies, (J- averaged values taken from NIST \citep{Palmer1977}) and the Considered Molecular Symmetries for the Treated Molecular States.}             
\label{table2}             
\begin{tabular}{ l l l l l l l }    % 7 columns

\hline \hline
\multirow{1}{*} {j} &\multirow{1}{*}  {Asymptotic atomic states} & \multirow{1}{*} {Asymptotic }& \multicolumn{4}{c} { Molecular  } \\
\multirow{1}{*}     &\multirow{1}{*}                                         & \multirow{1}{*} {energies (eV) }& \multicolumn{4}{c} {  symmetry} \\
 \hline
 1 & $\mathrm{Y}(4d^2(^3\mathrm{F})5s$ $\mathrm{a}^4\mathrm{ F})+\mathrm{H} (1s$ $^2\mathrm{S})$   & 1.35606353 & $^3\Sigma^-$  & $^3\Pi$  & $^3\Delta$  & $^3\Phi$  \\
 2 & $\mathrm{Y}(4d^2(^3\mathrm{P})5s$ $\mathrm{a}^4\mathrm{P})+\mathrm{H} (1s$ $^2\mathrm{S})$   & 1.88724197 & $^3\Sigma^-$  & $^3\Pi$  & & \\
 3 & $\mathrm{Y}(4d^2(^3\mathrm{F})5s$ $\mathrm{a}^2\mathrm{ F})+\mathrm{H} (1s$ $^2\mathrm{S})$  & 1.90027370 & $^3\Sigma^-$  & $^3\Pi$  & $^3\Delta$  & $^3\Phi$  \\
 4 & $\mathrm{Y}(4d5s(^3\mathrm{D})5p$ $\mathrm{z}^2\mathrm{D}^{\mathrm{o}})+\mathrm{H} (1s$ $^2\mathrm{S})$   & 1.99193807 & $^3\Sigma^-$  & $^3\Pi$  & $^3\Delta$  & \\
 5 & $\mathrm{Y}(4d5s(^3\mathrm{D}) 5p$ $\mathrm{z}^4\mathrm{D}^{\mathrm{o}})+\mathrm{H} (1s$ $^2\mathrm{S})$   & 2.03779275 &$^3\Sigma^-$  & $^3\Pi$  & $^3\Delta$  & \\
 6 & $\mathrm{Y}(4d^2(^3\mathrm{P})5s$ $\mathrm{a}^2\mathrm{P})+\mathrm{H} (1s$ $^2\mathrm{S})$   & 2.38515941 & $^3\Sigma^-$  & $^3\Pi$  & & \\
 7 & $\mathrm{Y}(4d5s(^1\mathrm{D})5p$ $\mathrm{y}^2\mathrm{D}^{\mathrm{o}})+\mathrm{H} (1s$ $^2\mathrm{S})$   & 2.99189369 & $^3\Sigma^-$  & $^3\Pi$  & $^3\Delta$  & \\
 8 & $\mathrm{Y}(4d^2(^3\mathrm{F})5p$ $\mathrm{z}^4\mathrm{G}^{\mathrm{o}})+\mathrm{H} (1s$ $^2\mathrm{S})$   & 3.55760755 & $^3\Sigma^-$  & $^3\Pi$  & $^3\Delta$  & $^3\Phi$  \\
 9 & $\mathrm{Y}(4d^3$ $\mathrm{b}^4\mathrm{ F})+\mathrm{H} (1s$ $^2\mathrm{S})$   & 3.62926285 & $^3\Sigma^-$  & $^3\Pi$  & $^3\Delta$  & $^3\Phi$  \\
 10 & $\mathrm{Y}(4d^3$ $\mathrm{b}^4\mathrm{P})+\mathrm{H} (1s$ $^2\mathrm{S})$   & 3.96471512 & $^3\Sigma^-$  & $^3\Pi$  & & \\
 11 & $\mathrm{Y}(4d^2(^3\mathrm{F})5p$ $\mathrm{y}^4\mathrm{D}^{\mathrm{o}})+\mathrm{H} (1s$ $^2\mathrm{S})$   & 4.11819020 & $^3\Sigma^-$  & $^3\Pi$  & $^3\Delta$  & \\
 12 & $\mathrm{Y}(4d^2(^3\mathrm{F})5p$ $\mathrm{z}^2\mathrm{G}^{\mathrm{o}})+\mathrm{H} (1s$ $^2\mathrm{S})$   & 4.14505113 & $^3\Sigma^-$  & $^3\Pi$  & $^3\Delta$  & $^3\Phi$  \\
 13 & $\mathrm{Y}(4d^3$ $^2\mathrm{P})+\mathrm{H} (1s$ $^2\mathrm{S})$   & 4.16751186 & $^3\Sigma^-$  & $^3\Pi$  & & \\
 14 & $\mathrm{Y}(4d^2(^3\mathrm{F})5p$ $\mathrm{x}^2\mathrm{D}^{\mathrm{o}})+\mathrm{H} (1s$ $^2\mathrm{S})$   & 4.20390258 & $^3\Sigma^-$  & $^3\Pi$  & $^3\Delta$  & \\
 15 & $\mathrm{Y}(5p^2(^3\mathrm{P})5s$ $\mathrm{e}^4\mathrm{P})+\mathrm{H} (1s$ $^2\mathrm{S})$   & 4.20451395 & $^3\Sigma^-$  & $^3\Pi$  & & \\
 16 & $\mathrm{Y}(4d^2(^3\mathrm{P})5p$ $\mathrm{z}^2\mathrm{ S}^{\mathrm{o}})+\mathrm{H} (1s$ $^2\mathrm{S})$   & 4.26979609 & $^3\Sigma^-$  & & & \\
 17 & $\mathrm{Y}(4d^2(^3\mathrm{P})5p$ $\mathrm{x}^4\mathrm{D}^{\mathrm{o}})+\mathrm{H} (1s$ $^2\mathrm{S})$   & 4.42392210 & $^3\Sigma^-$  & $^3\Pi$  & $^3\Delta$  & \\
 18 & $\mathrm{Y}(4d^2(^3\mathrm{P})5p$ $\mathrm{z}^4\mathrm{ S}^{\mathrm{o}})+\mathrm{H} (1s$ $^2\mathrm{S})$   & 4.48023931 &  $^3\Sigma^-$  & & & \\
 19 & $\mathrm{Y}(4d^2(^1\mathrm{D})5p$ $\mathrm{w}^2\mathrm{D}^{\mathrm{o}})+\mathrm{H} (1s$ $^2\mathrm{S})$   & 4.51950858 & $^3\Sigma^-$  & $^3\Pi$  & $^3\Delta$  & \\
 20 & $\mathrm{Y}(4d^{3}$ $^2\mathrm{F})+\mathrm{H} (1s$ $^2\mathrm{S})$   & 4.73081880 & $^3\Sigma^-$  & $^3\Pi$  & $^3\Delta$  & $^3\Phi$  \\
 21 & $\mathrm{Y}(4d5s(^3\mathrm{D})6p$ $^4\mathrm{D}^{\mathrm{o}})+\mathrm{H} (1s$ $^2\mathrm{S})$   & 4.77007840 & $^3\Sigma^-$  & $^3\Pi$  & $^3\Delta$  & \\
 22 & $\mathrm{Y}(4d^2(^1\mathrm{G})5p$ $\mathrm{y}^2 \mathrm{G}^{\mathrm{o}})+\mathrm{H} (1s$ $^2\mathrm{S})$   & 4.77079008 & $^3\Sigma^-$  & $^3\Pi$  & $^3\Delta$  & $^3\Phi$  \\
 23 & $\mathrm{Y}(4d5s(^3\mathrm{D})5d$ $^2\mathrm{P})+\mathrm{H} (1s$ $^2\mathrm{S})$   & 4.84620470 & $^3\Sigma^-$  & $^3\Pi$  &  &   \\
 24 & $\mathrm{Y}(4d5s(^3\mathrm{D})6p$ $^2\mathrm{D}^{\mathrm{o}})+\mathrm{H} (1s$ $^2\mathrm{S})$   & 4.86532200 & $^3\Sigma^-$  & $^3\Pi$  & $^3\Delta$  & \\
 25 & $\mathrm{Y}(4d5s(^3\mathrm{D})5d$ $^2\mathrm{ F})+\mathrm{H} (1s$ $^2\mathrm{S})$   & 4.89013405 & $^3\Sigma^-$  & $^3\Pi$  & $^3\Delta$  & $^3\Phi$  \\
 26 & $\mathrm{Y}(4d5s(^3\mathrm{D})5d$ $\mathrm{e}^4\mathrm{ F})+\mathrm{H} (1s$ $^2\mathrm{S})$   & 4.89072060 & $^3\Sigma^-$  & $^3\Pi$  & $^3\Delta$  & $^3\Phi$  \\
 27 & $\mathrm{Y}(4d5s(^3\mathrm{D})5d$ $\mathrm{f}^4\mathrm{P})+\mathrm{H} (1s$ $^2\mathrm{S})$   & 4.99504530 & $^3\Sigma^-$  & $^3\Pi$  & & \\
 28 & $\mathrm{Y}(4d^2(^3\mathrm{P})5p$ $\mathrm{v}^2 \mathrm{D}^{\mathrm{o}})+\mathrm{H} (1s$ $^2\mathrm{S})$   & 5.03828950 & $^3\Sigma^-$  & $^3\Pi$  & $^3\Delta$  & \\
 29 & $\mathrm{Y}(4d^2(^3\mathrm{ F})6s$ $^4\mathrm{ F})+\mathrm{H} (1s$ $^2\mathrm{S})$   & 5.22057470 & $^3\Sigma^-$  & $^3\Pi$  & $^3\Delta$  & $^3\Phi$  \\
 30 & $\mathrm{Y}(4d5s(^1\mathrm{D})5d$ $^2\mathrm{P})+\mathrm{H} (1s$ $^2\mathrm{S})$   & 5.28862380 & $^3\Sigma^-$  & $^3\Pi$  & & \\
 31 & $\mathrm{Y}(5p^2(^3\mathrm{P})4d$ $^4\mathrm{ F})+\mathrm{H} (1s$ $^2\mathrm{S})$   & 5.34319570 & $^3\Sigma^-$  & $^3\Pi$  & $^3\Delta$  & $^3\Phi$  \\
 32 & $\mathrm{Y}(4d5s(^1 \mathrm{D})5d$ $^2\mathrm{F})+\mathrm{H} (1s$ $^2\mathrm{S})$   & 5.44039490 & $^3\Sigma^-$  & $^3\Pi$  & $^3\Delta$  & $^3\Phi$  \\
 33 & $\mathrm{Y}(4d5s(^1 \mathrm{D})5d$ $\mathrm{ g}^4\mathrm{ F})+\mathrm{H} (1s$ $^2\mathrm{S})$   & 5.50070000 & $^3\Sigma^-$  & $^3\Pi$  & $^3\Delta$  & $^3\Phi$  \\
 34 & $\mathrm{Y}(5 p^2(^3\mathrm{P})4d$ $\mathrm{ h}^4 \mathrm{ F})+\mathrm{H} (1s$ $^2\mathrm{S})$   & 5.58788000 & $^3\Sigma^-$  & $^3\Pi$  & $^3\Delta$  & $^3\Phi$  \\
 35 & $\mathrm{Y}(5 p^2(^3\mathrm{P})4d$ $\mathrm{f}^2\mathrm{P})+\mathrm{H} (1s$ $^2\mathrm{S})$   & 5.69676460 & $^3\Sigma^-$  & $^3\Pi$  & & \\
 \hline
 36 & $\mathrm{Y}^{+}(4d^2$ $\mathrm{a}^3\mathrm{ F})+\mathrm{H}^{-}(1s^2$ $\mathrm{ S})$ & 6.45552120 & $^3\Sigma^-$  & $^3\Pi$  & $^3\Delta$  & $^3\Phi$  \\
\hline
\end{tabular}
\end{table}

The ion-molecular state Y$^+$+H$^-$ encompasses different electronic configurations, such as: the ground state Y$^+(5s^2$ $\mathrm{a}^1\mathrm{S})$+H$^-(1s^2$ $^1\mathrm{S})$  only includes one molecular symmetry $^1\Sigma^+$, the first excited state Y$^+(4d5s$ $\mathrm{a}^3\mathrm{D})$+H$^-(1s^{2}$ $^1\mathrm{S})$ includes 3 molecular symmetries: $^3 \Sigma^{+}$, $^3\Pi$, and $^3 \Delta$ and the third excited state Y$^+(4d^2$ $\mathrm{a}^3\mathrm{F})$+H$^{-}(1 s^2$ $^1\mathrm{S})$ includes 4 molecular symmetries: $^3 \Sigma^{-}$, $^3\Pi$, $^3\Delta$, and $^3\Phi$. In our calculations, only covalent scattering channels which have the same molecular symmetries with the ion-molecular states are considered. Due to a fact that there are two symmetries for $\Sigma$ state, $\Sigma^{+}$ and $\Sigma^{-}$, so the calculations for this symmetry are divided into two groups. The first group is the single-electron transfer process which involving the interaction between the covalent state  $\mathrm{Y}[(4d^2\mathrm{n}l) /(4d5s\mathrm{n}l) /(5s^2 \mathrm{n}l) /(5 p^2 \mathrm{n} l)$ $^{2,4}L]+\mathrm{H}$ and the ion state $\mathrm{Y}^+[(5s^2$ $\mathrm{a}^1\mathrm{S})/(4d5s$ $\mathrm{a}^3\mathrm{D})]+\mathrm{H}^-$. The asymptotic atomic states and corresponding energies of the related molecular states are shown in Table \ref{table1}. The second group includes the single-electron transfer process which involving the interaction between the covalent state $\mathrm{Y}[(4d^2 \mathrm{n}l) /(4d5s\mathrm{n}l) /(5 p^2 \mathrm{n}l)$ $^{2,4}L]+\mathrm{H}$ and the ion state $\mathrm{Y}^{+}(4 d^2$ $\mathrm{a}^3\mathrm{F})+\mathrm{H}^{-}$. The asymptotic atomic states and corresponding energies of the related molecular states are displayed in Table \ref{table2}. Considering that the initial and final scattering channels in the same process may include different molecular symmetries, the rate coefficients of these non-elastic processes need to be computed independently for each molecular symmetry, and then summed to obtain the total rate coefficient of the corresponding process.

\begin{figure}
\centering 
	\includegraphics[width=8.7cm]{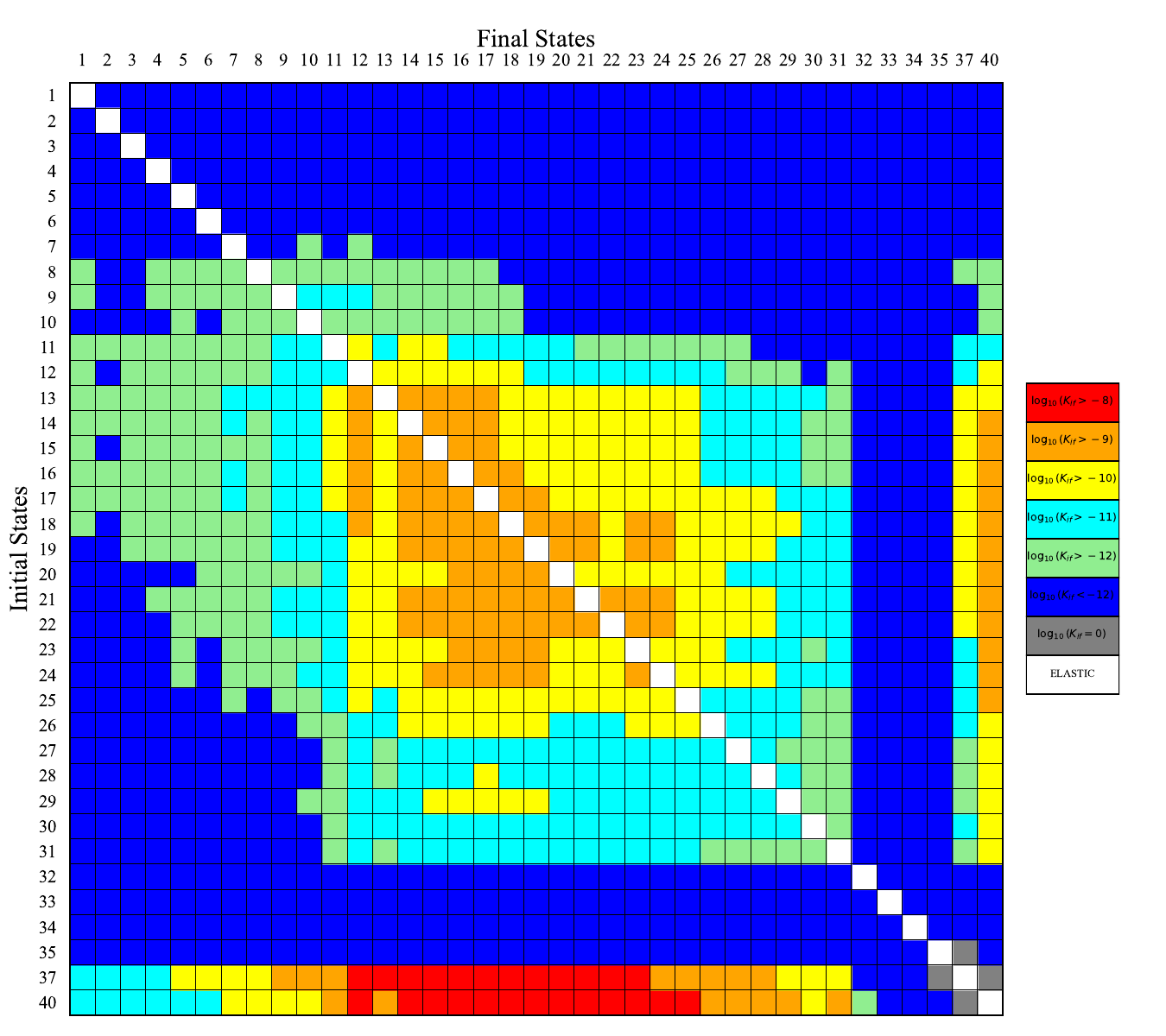} %width=\columnwidth
    \caption{ Rate coefficients (in cm$^3s^{-1}$) for the partial excitation, de-excitation, mutual neutralization, and ion-pair formation processes in Y$^+(5s^2$ $\mathrm{a} ^1\mathrm{S})$+H$^-$ and Y$^+(4d5s$ $\mathrm{a} ^3\mathrm{D})$+H$^-$ collisions at temperature of T= 6000K. Initial and final state (or scattering channels) labels are presented in Table \ref{table1}.}
    \label{fig1}
\end{figure}

\begin{figure}
\centering 
	\includegraphics[width=8.7cm]{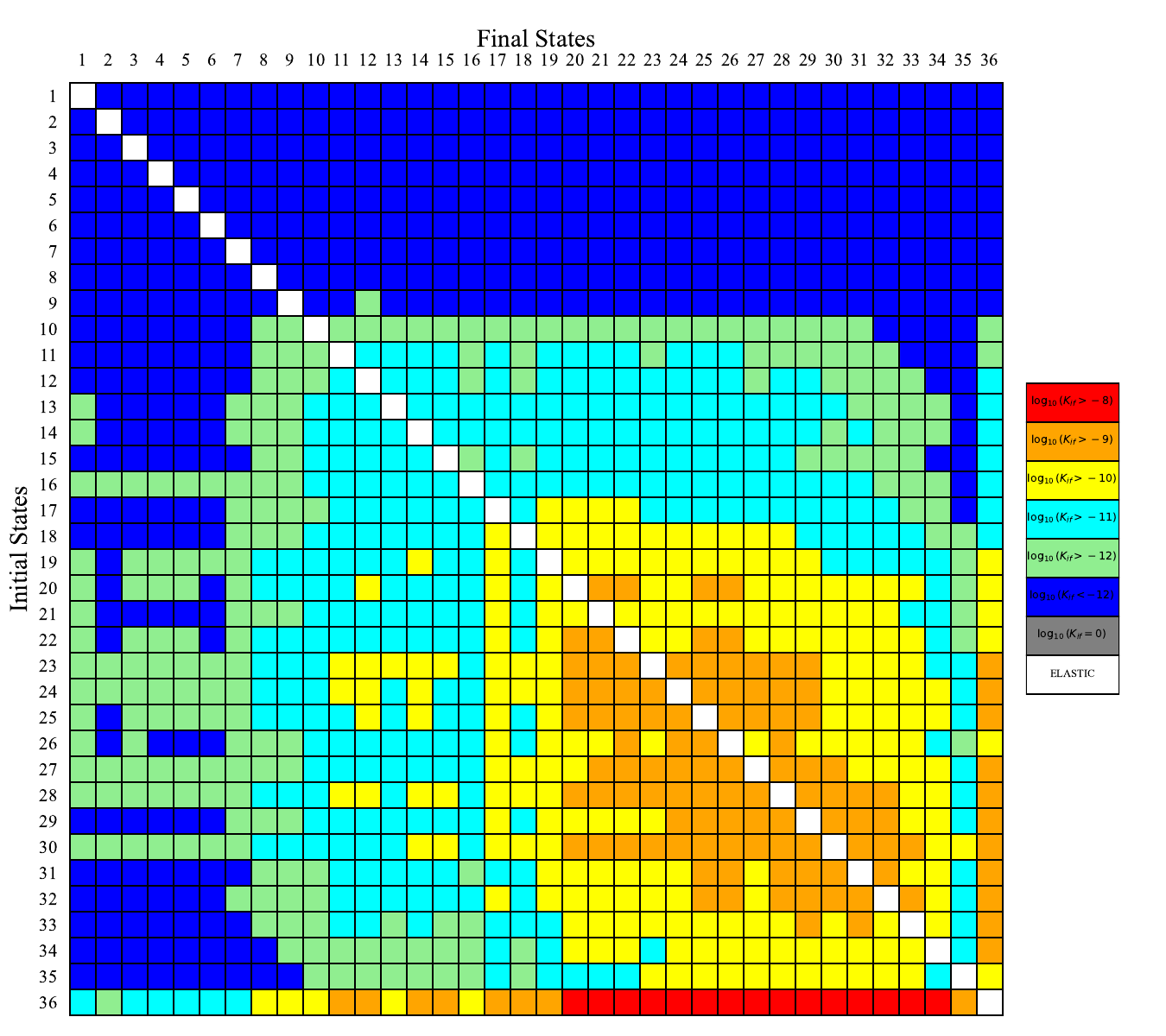}
    \caption{Rate coefficients (in cm$^3s^{-1}$) for the partial excitation, de-excitation, mutual neutralization, and ion-pair formation processes in Y$^+(4d^2$ $\mathrm{a}^3\mathrm{F})$+H$^-$ collisions at temperature of T= 6000K. Initial and final state (or scattering channels) labels are presented in Table \ref{table2}.}
    \label{fig2}
\end{figure}

We calculate the rate coefficients for the excitation, de-excitation, ion pair formation, and mutual neutralization processes for the collisions of $\mathrm{Y}+\mathrm{H}$ and $\mathrm{Y}^{+}+\mathrm{H}^{-}$ at temperatures ranging from 1000 to 10000K by employing the simplified model. In Fig.\ref{fig1}, we present the results of rate coefficients for non-elastic processes in collisions of $\mathrm{Y}[(4d^2\mathrm{n}l) /(4d5s\mathrm{n}l) /(5s^2 \mathrm{n}l) /(5 p^2 \mathrm{n} l)$ $^{2,4}L]+\mathrm{H}$ and $\mathrm{Y}^+[(5s^2$ $\mathrm{a}^1\mathrm{S})/(4d5s$ $\mathrm{a}^3\mathrm{D})]+\mathrm{H}^-$ at a temperature of T=6000K. The complete data for rate coefficients for the temperatures ranging from 1000 to 10000K can be seen in the supplementary materials. In the present calculations, the processes of $\mathrm{Y}^+(5s^2$ $\mathrm{a}^1\mathrm{S})+\mathrm{H}^{-}$$\rightarrow$$\mathrm{Y}[(4d^2\mathrm{n}l) /(4d5s\mathrm{n}l) /(5s^2 \mathrm{n}l) /(5 p^2 \mathrm{n} l)$ $^{2,4}L]$ and $\mathrm{Y}^+(4d5s$ $\mathrm{a}^3\mathrm{D})+\mathrm{H}^-$$\rightarrow$$\mathrm{Y}[(4d^2\mathrm{n}l) /(4d5s\mathrm{n}l) /(5s^2 \mathrm{n}l) /(5 p^2 \mathrm{n} l)$ $^{2,4}L]$ are individually considered due to the distinct spins (or multiplicities) of the associated molecular states. As mentioned above, if the associated molecular states corresponding to the initial and final scattering channels have different spins (or multiplicities), the summation of the rate coefficient from these states yields the total rate coefficient of the corresponding collision processes.

It can be seen from Fig. \ref{fig1} that the largest rate coefficients are corresponding to the mutual neutralization process. Specifically, for the mutual neutralization process of $\mathrm{Y}^+(5s^2$ $\mathrm{a}^1\mathrm{S})+\mathrm{H}^{-}$, the final scattering channels with rate coefficients exceeding $10^{-8} \mathrm{ cm}^3 \mathrm{ s}^{-1}$ are labeled as $j$=12-23, and the corresponding specific states of the yttrium atom are displayed in Table \ref{table1}. The rate coefficient for the scattering channel which denoted by $j$=17 (Y$(5s^2(^2\mathrm{D})5d$ $e^2$D)+H) demonstrates a peak value of $6.91\times 10^{-8}\mathrm{cm}^3\mathrm{s}^{-1}$. Within the mutual neutralization process of $\mathrm{Y}^+(4d5s$ $\mathrm{a}^3\mathrm{D})+\mathrm{H}^-$, the final channels with rate coefficients exceeding $10^{-8} \mathrm{ cm}^3 \mathrm{ s}^{-1}$ are labeled as$j$=12, 14-25. The specific states corresponding to these channels can been seen Table \ref{table1}. It should be noted that the rate coefficient of the scattering channels which marked by $j$=17 (Y$(5s^2(^2\mathrm{D})5d$ $e^2$D)+H) also show a peak value of $6.77\times 10^{-8}\mathrm{cm}^3\mathrm{s}^{-1}$. 

Figure \ref{fig2}  presents the rate coefficients for non-elastic processes in collisions of $\mathrm{Y}[(4d^2 \mathrm{n}l) /(4d5s\mathrm{n}l) /(5p^2 \mathrm{n}l)$ $^{2,4}L]+\mathrm{H}$ and $\mathrm{Y}^{+}(4 d^2$ $\mathrm{a}^3\mathrm{F})+\mathrm{H}^{-}$  at a temperature of T=6000K. In the mutual neutralization process of $\mathrm{Y}^{+}(4 d^2$ $\mathrm{a}^3\mathrm{F})+\mathrm{H}^{-}$ system, the final scattering channels with rate coefficients exceeding $10^{-8} \mathrm{ cm}^3 \mathrm{ s}^{-1}$ are denoted as $j$=20-34, and the corresponding specific states of the yttrium atom are displayed in Table \ref{table2}. The rate coefficients of the final scattering channel which marked by $j$=29 (Y$(4d^2(^3\mathrm{F})6s$ $^4$F)+H) show a peak value of $6.95\times 10^{-8}\mathrm{cm}^3\mathrm{s}^{-1}$. The other rate coefficients with intermediate values ($10^{-12}$-$10^{-8} \mathrm{ cm}^3 \mathrm{ s}^{-1}$) which represented by green, cyan, yellow and orange squares correspond to the neutralization, ion pair formation, excitation, and de-excitation processes, respectively. It is to be noted that the maximum rate coefficients for the excitation and de-excitation processes are an order of magnitude smaller than those of the mutual neutralization process. The rate coefficients for the elastic processes are denoted by the white squares on the diagonals in Figs \ref{fig1} and \ref{fig2}.

It should be noted that the rate coefficients with intermediate values of $10^{-12}$-$10^{-8} \mathrm{ cm}^3 \mathrm{ s}^{-1}$ in Fig. \ref{fig2}, correspond to the transitions between higher excited states of the collision systems. \citet{Belyaev2017a} clarified that the simplified model produces the intermediate-to-high values of rate coefficients for the processes which involving atomic states whose asymptotic energies approach the positions of the "optimal windows".  The "optical windows" means a range of atomic binding energy, and the scattering channels which asymptotic energies near these positions will have larger rate coefficients. These "optimal windows" are approximately 2eV lower than the ionization energy of Y atoms. In the case of the ground state $\mathrm{Y}^+(5s^2$ $\mathrm{a}^1\mathrm{S})$ and the first excited state $\mathrm{Y}^{+}(4d5s$ $\mathrm{a}^3\mathrm{D})$, the positions of optimal window are around 4.21eV and 4.31eV (excitation energy of Y atoms), respectively. Whereas for the third excited state of $\mathrm{Y}^{+}(4d^2$ $\mathrm{a}^3\mathrm{ F})$, the position of optimal window is approximately at 5.21eV. This accounts for the maximum values of the rate coefficient obtained for the scattering channels of $j$=17 and 29 in the neutralization processes in the first group (Y$^+(5s^2$ $\mathrm{a} ^1\mathrm{S})$+H$^-$ and Y$^+(4d5s$ $\mathrm{a} ^3\mathrm{D})$+H$^-$)  and second group (Y$^+(4d^2$ $\mathrm{a}^3\mathrm{F})$+H$^-$), respectively. Thus, for different ionic scattering channels, the “optimal window” selects different final scattering channels \citep{Voronov2022}.

\begin{figure*}
    \centering 
    \includegraphics[width=16cm]{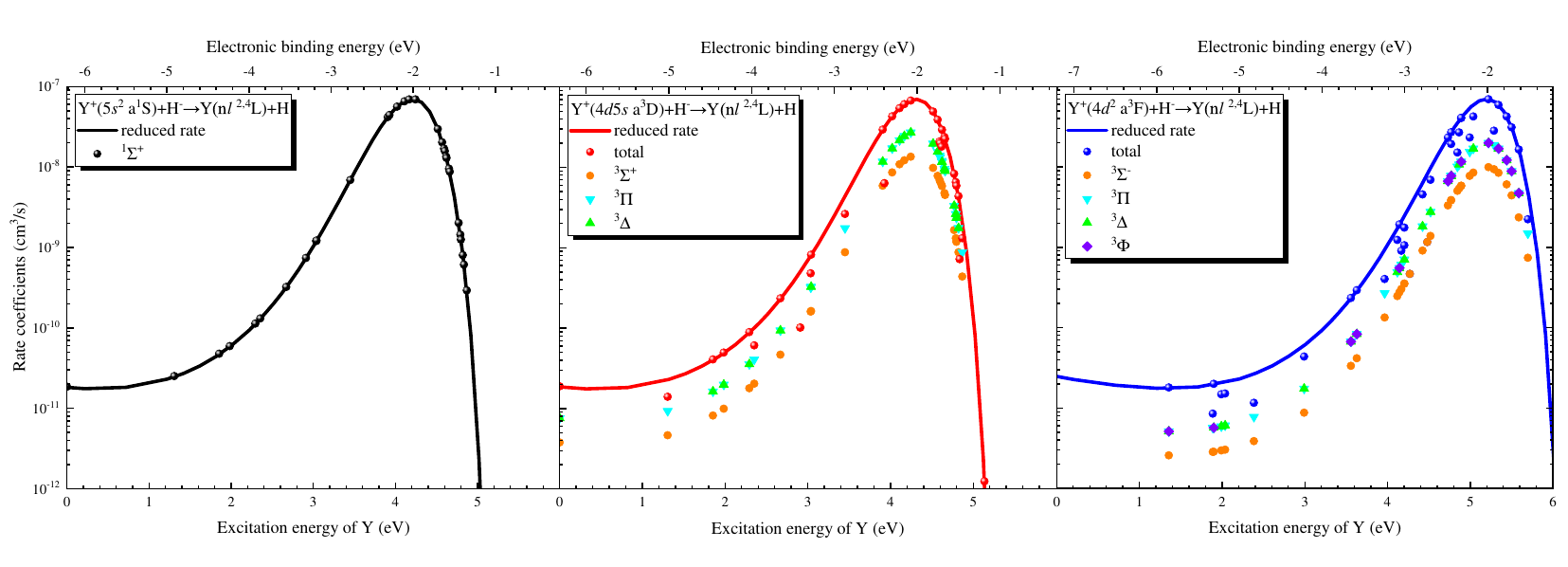}
    \caption{Rate coefficients for mutual neutralization in yttrium–hydrogen collisions (symbols) at T=6000K as a function of the excitation energy of Y in the final state. Note the different reduced rate coefficient dependent (solid lines) for different initial ionic states Y$^+$+H$^-$.}
    \label{fig3}
\end{figure*}

Figure \ref{fig3}  presents the variation of the total rate coefficient of three sets of mutual neutralization processes at a temperature of T=6000K. The left panel shows the variation of the rate coefficient with the excitation energy of yttrium atoms for the collisions with ground ionic state  $\mathrm{Y}^{+}(5s^2$ $\mathrm{a}^1\mathrm{S})+\mathrm{H}^-$, which characterized by a single molecular symmetry ($^1\Sigma^+$). The middle panel of the figure describes the results for the collisions between $\mathrm{H}^{-}$ ions and the first excited ionic state of $\mathrm{Y}^{+}(4d5s$ $\mathrm{a}^3\mathrm{D})$, and for this collision system the molecule [YH] exhibiting three symmetries ($^3\Sigma^+$, $^3\Pi$ and $^3\Delta$), while the right panel displays the results for the collisions of $\mathrm{H}^{-}$ ions with the third excited ionic state of $\mathrm{Y}^{+}(4d^2$ $\mathrm{a}^3\mathrm{F})$, with four molecular symmetries ($^3\Sigma^-$, $^3\Pi$, $^3\Delta$ and $^3\Phi$). In this figure, solid lines represent the reduced rate coefficients. Under the simplified model, non-adiabatic nuclear dynamics remain the same for each molecular symmetry, and the reduced rate coefficients also remain consistent for each ionic scattering channel. The difference between the reduced rate coefficients and the rate coefficients calculated in this study originates from the fact that some covalent states do not have all the molecular symmetries like those of the ionic states. When all the molecular symmetries of the initial ionic scattering channel are all included into the final covalent scattering channels in the calculation, the rate coefficients are equal to the reduced rate coefficients. On the contrary, if not all molecular symmetries are included, the rate coefficients are determined by the statistical probability of the contributed molecular symmetries multiplied by the reduced rate coefficients, resulting in the lower values of the rate coefficients corresponding to the reduced rate coefficients. This can be observed in Fig. \ref{fig3}. Furthermore, this figure also illustrates that the "optimal window" is typically situated near the point where the electron binding energy equals -2eV during atom-hydrogen collisions.

\begin{table} 
\setlength{\tabcolsep}{2pt}
 \renewcommand{\arraystretch}{0.7} % 
\centering   
\caption{YH$^+$ molecular states for yttrium ionic configuration Y$^{2+}(4p^64d$ $^2$D),  the corresponding asymptotic atomic states (scattering channels), the asymptotic energies,  ($J$- averaged values taken from NIST  \citep{Nilsson_1991}) and the considered molecular symmetries for the treated molecular states.}             
\label{table3}             
\begin{tabular}{ l l l l l l l }
\hline\hline
\multirow{1}{*} {j} &\multirow{1}{*}  {Asymptotic atomic states} & \multirow{1}{*} {Asymptotic }& \multicolumn{4}{c} { Molecular  } \\
\multirow{1}{*}     &\multirow{1}{*}                                         & \multirow{1}{*} {energies (eV) }& \multicolumn{4}{c} {  symmetry} \\
 \hline
 1 & $\mathrm{Y}^{+} (5s^2$ $\mathrm{a}^1 \mathrm{S} )+\mathrm{H} (1s$ $^2 \mathrm{S} )$ & 0.0000000 & $^2\Sigma^+$  &  &  &   \\
  2 & $\mathrm{Y}^{+} (4d5s$ $\mathrm{a}^3 \mathrm{D} )+\mathrm{H} (1s$ $^2 \mathrm{S} )$ & 0.1041713 & $^2\Sigma^+$& & $^2\Pi$  & $^2\Delta$ \\
  3 & $\mathrm{Y}^{+} (4d5s$ $\mathrm{a}^1 \mathrm{D} )+\mathrm{H} (1s$ $^2\mathrm{S} )$ & 0.4086742 & $^2\Sigma^+$& & $^2\Pi$  & $^2\Delta$ \\
  4 & $\mathrm{Y}^{+} (4d^2$ $\mathrm{a}^3 \mathrm{F} )+\mathrm{H} (1s$ $^2 \mathrm{S} )$ & 0.9922612 & & $^2\Sigma^-$ & $^2\Pi$  & $^2\Delta$ \\
  5 & $\mathrm{Y}^{+} (4d^2$ $\mathrm{a}^3 \mathrm{P} )+\mathrm{H} (1s$ $^2 \mathrm{S} )$ & 1.7213196 & &  $^2\Sigma^-$ & $^2\Pi$ & \\
  6 & $\mathrm{Y}^{+} (4d^2$ $\mathrm{b}^1 \mathrm{D} )+\mathrm{H} (1s$ $^2 \mathrm{S} )$ & 1.8390405 & $^2\Sigma^+$& & $^2\Pi$  & $^2\Delta$ \\
  7 & $\mathrm{Y}^{+} (4d^2$ $\mathrm{a}^1\mathrm{G} )+\mathrm{H} (1s$ $^2 \mathrm{S} )$ & 1.9444324 & $^2\Sigma^+$& & $^2\Pi$  & $^2\Delta$ \\
  8 & $\mathrm{Y}^{+} (4d5p$ $\mathrm{z}^3\mathrm{P}^{\mathrm{o}})+\mathrm{H} (1s$ $^2 \mathrm{S} )$ & 2.9068173 & $^2\Sigma^+$ & & $^2\Pi$ & \\
  9 & $\mathrm{Y}^{+} (4d^2$ $^1\mathrm{S})+\mathrm{H} ( 1s$ $^2 \mathrm{S} )$ & 3.1083155 & $^2\Sigma^+$ & & & \\
  10 & $\mathrm{Y}^{+}(4d5p$ $\mathrm{z}^1\mathrm{D}^{\mathrm{o}} )+\mathrm{H} ( 1s$ $^2 \mathrm{S} )$ & 3.2418461 & & $^2\Sigma^-$ & $^2\Pi$  & $^2\Delta$ \\
  11 & $\mathrm{Y}^{+}(4d5p$ $\mathrm{z}^3\mathrm{F}^{\mathrm{o}} )+\mathrm{H} ( 1s$ $^2 \mathrm{S} )$ & 3.3757211 & $^2\Sigma^+$& & $^2\Pi$  & $^2\Delta$ \\
  12 & $\mathrm{Y}^{+}(4d5p$ $\mathrm{z}^{\mathrm{1}} \mathrm{P}^{\mathrm{o}} )+\mathrm{H} ( 1s$ $^2 \mathrm{S} )$ & 3.4116359 & $^2\Sigma^+$ & & $^2\Pi$  & \\
  13 & $\mathrm{Y}^{+}(4d5p$ $\mathrm{z}^3\mathrm{D}^{\mathrm{o}} )+\mathrm{H} ( 1s$ $^2 \mathrm{S} )$ & 3.5453635 & & $^2\Sigma^-$ & $^2\Pi$  & $^2\Delta$ \\
  14 & $\mathrm{Y}^{+}(4d5p$ $\mathrm{y}^3\mathrm{P}^{\mathrm{o}} )+\mathrm{H} ( 1s$ $^2 \mathrm{S} )$ & 3.9735433 & $^2\Sigma^-$  &   & $^2\Pi$ & \\
  15 & $\mathrm{Y}^{+}(4d5p$ $\mathrm{z}^1\mathrm{F}^{\mathrm{o}} )+\mathrm{H} ( 1s$ $^2 \mathrm{S} )$ & 4.1332274 &$^2\Sigma^+$& & $^2\Pi$  & $^2\Delta$  \\
  16 & $\mathrm{Y}^{+}(4d5p$ $\mathrm{y}^1\mathrm{P}^{\mathrm{o}} )+\mathrm{H} ( 1s$ $^2 \mathrm{S} )$ & 5.5258071 & $^2\Sigma^-$  & & $^2\Pi$ & \\
  17 & $\mathrm{Y}^{+}(4d6s$ $\mathrm{e}^3\mathrm{D} )+\mathrm{H} ( 1s$ $^2 \mathrm{S} )$ & 6.8136859 & $^2\Sigma^-$ & & $^2\Pi$ & $^2\Delta$ \\
  18 & $\mathrm{Y}^{+}(4d6s$ $\mathrm{e}^{\mathrm{l}} \mathrm{D} )+\mathrm{H} ( 1s$ $^2 \mathrm{S} )$ & 6.9090842 &$^2\Sigma^+$& & $^2\Pi$  & $^2\Delta$  \\
  19 & $\mathrm{Y}^{+}(4d6s$ $\mathrm{e}^3\mathrm{S} )+\mathrm{H} ( 1s$ $^2 \mathrm{S} )$ & 7.2237209 &$^2\Sigma^-$  & & & \\
  20 & $\mathrm{Y}^{+}(4d5d$ $\mathrm{e}^1\mathrm{F} )+\mathrm{H} ( 1s$ $^2 \mathrm{S} )$ & 7.2572038 & & $^2\Sigma^-$ & $^2\Pi$  & $^2\Delta$ \\
  21 & $\mathrm{Y}^{+}(4d5d$ ${\mathrm{f}}^3\mathrm{D} )+\mathrm{H} ( 1s$ $^2 \mathrm{S} )$ & 7.2803995 &$^2\Sigma^+$& & $^2\Pi$  & $^2\Delta$  \\
  22 & $\mathrm{Y}^{+}(5p^2$ $\mathrm{e}^3 \mathrm{P} )+\mathrm{H} ( 1s$ $^2 \mathrm{S} )$ & 7.2873479 & & $^2\Sigma^-$ & $^2\Pi$ & \\
  23 & $\mathrm{Y}^{+}(4d5d$ $\mathrm{e}^3 \mathrm{G} )+\mathrm{H} ( 1s$ $^2 \mathrm{S} )$ & 7.3373339 &$^2\Sigma^+$& & $^2\Pi$  & $^2\Delta$  \\
  24 & $\mathrm{Y}^{+}(4d5d$ $\mathrm{e}^1 \mathrm{P} )+\mathrm{H} ( 1s$ $^2 \mathrm{S} )$ & 7.4039449 & & $^2\Sigma^-$ & $^2\Pi$ & \\
  25 & $\mathrm{Y}^{+}(5p^2$ $\mathrm{f}^1 \mathrm{D} )+\mathrm{H} ( 1s$ $^2 S )$ & 7.5054978 &$^2\Sigma^+$& & $^2\Pi$  & $^2\Delta$  \\
  26 & $\mathrm{Y}^{+}(4d5d$ $\mathrm{f}^3 \mathrm{S} )+\mathrm{H} ( 1s$ $^2 \mathrm{S} )$ & 7.5880390 & $^2\Sigma^-$  & & & \\
  27 & $\mathrm{Y}^{+}(4d5d$ $\mathrm{e}^3 \mathrm{F})+\mathrm{H} ( 1s$ $^2 \mathrm{S} )$ & 7.6049447& & $^2\Sigma^-$ & $^2\Pi$  & $^2\Delta$\\
  28 & $\mathrm{Y}^{+}(4d6s$ $\mathrm{f}^1 \mathrm{S})+\mathrm{H} ( 1s$ $^2 \mathrm{S} )$ & 7.6088181 & $^2\Sigma^-$  & & & \\
  29 & $\mathrm{Y}^{+}(4d5d$ $\mathrm{g}^1 \mathrm{D})+\mathrm{H} ( 1s$ $^2 \mathrm{S} )$ & 7.7487058 &$^2\Sigma^+$& & $^2\Pi$  & $^2\Delta$  \\
  30 & $\mathrm{Y}^{+}(4d6p$ $^1\mathrm{D}^{\mathrm{o}})+\mathrm{H} ( 1s$ $^2 \mathrm{S} )$ & 7.8517126 & & $^2\Sigma^-$ & $^2\Pi$  & $^2\Delta$ \\
  31 & $\mathrm{Y}^{+}(4d5d$ $\mathrm{e}^1 \mathrm{G} )+\mathrm{H} ( 1s$ $^2 \mathrm{S} )$ & 7.8544279 &$^2\Sigma^+$& & $^2\Pi$  & $^2\Delta$  \\
  32 & $\mathrm{Y}^{+}(4d6p$ $^3\mathrm{D}^{\mathrm{o}})+\mathrm{H} ( 1s$ $^2 \mathrm{S} )$ & 7.8821060 & & $^2\Sigma^-$ & $^2\Pi$  & $^2\Delta$ \\
  33 & $\mathrm{Y}^{+}(4d6p$ $^3\mathrm{F}^{\mathrm{o}})+\mathrm{H} ( 1s$ $^2 \mathrm{S} )$ & 7.9033562 &$^2\Sigma^+$& & $^2\Pi$  & $^2\Delta$  \\
  34 & $\mathrm{Y}^{+}(4d5d$ $\mathrm{f}^3\mathrm{P} )+\mathrm{H} ( 1s$ $^2 \mathrm{S} )$ & 7.9388951 & & $^2\Sigma^-$ & $^2\Pi$ & \\
  35 & $\mathrm{Y}^{+}(4d6p$ $^3\mathrm{P}^{\mathrm{o}})+\mathrm{H} ( 1s$ $^2 \mathrm{S} )$ & 8.0079394 & $^2\Sigma^+$  & & $^2\Pi$ & \\
  36 & $\mathrm{Y}^{+}(5s5d$ $\mathrm{g}^3 \mathrm{D} )+\mathrm{H} ( 1s$ $^2 \mathrm{S} )$ & 8.0754167 &$^2\Sigma^+$& & $^2\Pi$  & $^2\Delta$  \\
  37 & $\mathrm{Y}^{+}(4d6p$ $^1\mathrm{F}^{\mathrm{o}})+\mathrm{H} ( 1s$ $^2 \mathrm{S} )$ & 8.0850862 &$^2\Sigma^+$& & $^2\Pi$  & $^2\Delta$  \\
  38 & $\mathrm{Y}^{+}(4d6p$ $^1\mathrm{P}^{\mathrm{o}})+\mathrm{H} ( 1s$ $^2 \mathrm{S} )$ & 8.1166115 & $^2\Sigma^+$ & & $^2\Pi$ & \\
  39 & $\mathrm{Y}^{+}(4d5d$ $^1\mathrm{S} )+\mathrm{H} ( 1s$ $^2 \mathrm{S} )$ & 8.1188332 & $^2\Sigma^+$ & & & \\
  40 & $\mathrm{Y}^{+}(4d6p$ $^3\mathrm{P}^{\mathrm{o}})+\mathrm{H} ( 1s$ $^2 \mathrm{S} )$ & 8.5676363 & $^2\Sigma^+$ & & $^2\Pi$ & \\
  41 & $\mathrm{Y}^{+}(4d4f$ $^1\mathrm{G}^{\mathrm{o}})+\mathrm{H} ( 1s$ $^2 \mathrm{S} )$ & 8.6474561 & & $^2\Sigma^-$ & $^2\Pi$  & $^2\Delta$ \\
  42 & $\mathrm{Y}^{+}(4d4f$ $^3\mathrm{F}^{\mathrm{o}} )+\mathrm{H} ( 1s$ $^2 \mathrm{S} )$ & 8.6666397 &$^2\Sigma^+$& & $^2\Pi$  & $^2\Delta$  \\
  43 & $\mathrm{Y}^{+}(4d4f$ $^3\mathrm{H}^{\mathrm{o}})+\mathrm{H} ( 1s$ $^2 \mathrm{S} )$ & 8.6890561 &$^2\Sigma^+$& & $^2\Pi$  & $^2\Delta$  \\
  44 & $\mathrm{Y}^{+}(5s5d$ $\mathrm{h}^1\mathrm{D} )+\mathrm{H} ( 1s$ $^2 \mathrm{S} )$ & 8.7066790 &$^2\Sigma^+$& & $^2\Pi$  & $^2\Delta$  \\
  45 & $\mathrm{Y}^{+}(4d4f$ $^3\mathrm{G}^{\mathrm{o}})+\mathrm{H} ( 1s$ $^2 \mathrm{S} )$ & 8.7120752 & & $^2\Sigma^-$ & $^2\Pi$  & $^2\Delta$\\
  46 & $\mathrm{Y}^{+}(4d4f$ $^1\mathrm{D}^{\mathrm{o}})+\mathrm{H} ( 1s$ $^2 \mathrm{S} )$ & 8.7525910 & & $^2\Sigma^-$ & $^2\Pi$  & $^2\Delta$ \\
  47 & $\mathrm{Y}^{+}(5s6p$ $^{1}\mathrm{P}^{\circ} )+\mathrm{H} ( 1s$ $^2 \mathrm{S} )$ & 8.7557621 & $^2\Sigma^+$ & & $^2\Pi$& \\
  48 & $\mathrm{Y}^{+}(4d4f$ $^3\mathrm{D}^{\mathrm{o}})+\mathrm{H} ( 1s$ $^2 \mathrm{S} )$ & 8.7837765 & & $^2\Sigma^-$ & $^2\Pi$  & $^2\Delta$ \\
  49 & $\mathrm{Y}^{+}(4d4f$ $^1\mathrm{F}^{\mathrm{o}})+\mathrm{H} ( 1s$ $^2 \mathrm{S} )$ & 8.8250075 &$^2\Sigma^+$& & $^2\Pi$  & $^2\Delta$  \\
  50 & $\mathrm{Y}^{+}(4d4f$ $^3\mathrm{P}^{\mathrm{o}})+\mathrm{H} ( 1s$ $^2 \mathrm{S} )$ & 8.8585046 & $^2\Sigma^+$ & & $^2\Pi$ & \\
  51 & $\mathrm{Y}^{+}(4d4f$ $^1\mathrm{H}^{\mathrm{o}})+\mathrm{H} ( 1s$ $^2 \mathrm{S} )$ & 8.8816857 &$^2\Sigma^+$& & $^2\Pi$  & $^2\Delta$  \\
  52 & $\mathrm{Y}^{+}(4d4f$ $^1\mathrm{P}^{\mathrm{o}})+\mathrm{H} ( 1s$ $^2 \mathrm{S} )$ & 8.9367566 & $^2\Sigma^+$ & & $^2\Pi$ & \\
  53 & $\mathrm{Y}^{+}(4d7s$ $^3\mathrm{D} )+\mathrm{H} ( 1s$ $^2 \mathrm{S} )$ & 9.1335017 &$^2\Sigma^+$& & $^2\Pi$  & $^2\Delta$ \\
  54 & $\mathrm{Y}^{+}(4d7s$ $^1\mathrm{D} )+\mathrm{H} ( 1s$ $^2 \mathrm{S} )$ & 9.2470592 &$^2\Sigma^+$& & $^2\Pi$  & $^2\Delta$ \\
  55 & $\mathrm{Y}^{+}(4d6d$ $^1\mathrm{F} )+\mathrm{H} ( 1s$ $^2 \mathrm{S} )$ & 9.3613014 & & $^2\Sigma^-$ & $^2\Pi$  & $^2\Delta$\\
  56 & $\mathrm{Y}^{+}(4d6d$ $^3\mathrm{D} )+\mathrm{H} ( 1s$ $^2\mathrm{S} )$ & 9.3654408 &$^2\Sigma^+$& & $^2\Pi$  & $^2\Delta$  \\
   57 & $\mathrm{Y}^{+}(4d6d$ $^3\mathrm{G} )+\mathrm{H} ( 1s$ $^2 \mathrm{S} )$ & 9.3927240 &$^2\Sigma^+$& & $^2\Pi$  & $^2\Delta$ \\
  \hline
  \end{tabular}
  \end{table}

 \begin{table} 
\setlength{\tabcolsep}{2pt}
 \renewcommand{\arraystretch}{0.7} 
\centering   
\caption*{ Continue }                      
\begin{tabular}{ l l l l l l l }
\hline\hline
\multirow{1}{*} {j} &\multirow{1}{*}  {Asymptotic atomic states} & \multirow{1}{*} {Asymptotic }& \multicolumn{4}{c} { Molecular  } \\
\multirow{1}{*}     &\multirow{1}{*}                                         & \multirow{1}{*} {energies (eV) }& \multicolumn{4}{c} {  symmetry} \\
 \hline
  58 & $\mathrm{Y}^{+}(5s4f$ $^3\mathrm{F}^{\mathrm{o}})+\mathrm{H} ( 1s$ $^2 \mathrm{S} )$ & 9.4258129 &$^2\Sigma^+$& & $^2\Pi$  & $^2\Delta$  \\
  59 & $\mathrm{Y}^{+}(4d6d$ $^3\mathrm{S} )+\mathrm{H} ( 1s$ $^2 \mathrm{S} )$ & 9.4311710 & $^2\Sigma^+$ & & & \\
   60 & $\mathrm{Y}^{+}(5s4f$ $^1\mathrm{F}^{\mathrm{o}})+\mathrm{H} ( 1s$ $^2 \mathrm{S} )$ & 9.4631361 &$^2\Sigma^+$& & $^2\Pi$  & $^2\Delta$ \\
  61 & $\mathrm{Y}^{+}(4d6d$ $^1\mathrm{P})+\mathrm{H} ( 1s$ $^2 \mathrm{S} )$ & 9.4856132 & & $^2\Sigma^-$ & $^2\Pi$ & \\
  62 & $\mathrm{Y}^{+}(4d6d$ $^3\mathrm{F})+\mathrm{H} ( 1s$ $^2 \mathrm{S} )$ & 9.4870896 & & $^2\Sigma^-$ & $^2\Pi$  & $^2\Delta$\\
  63 & $\mathrm{Y}^{+}(4d6d$ $^{\mathrm{l}} \mathrm{D} )+\mathrm{H} ( 1s$ $^2 \mathrm{S} )$ & 9.5979664 &$^2\Sigma^+$& & $^2\Pi$  & $^2\Delta$ \\
   64 & $\mathrm{Y}^{+}(4d7p$ $^1\mathrm{D}^{\mathrm{o}})+\mathrm{H} ( 1s$ $^2 \mathrm{S} )$ & 9.6000624 & & $^2\Sigma^-$ & $^2\Pi$  & $^2\Delta$\\
  65 & $\mathrm{Y}^{+}(4d6d$ $^3 \mathrm{P} )+\mathrm{H} ( 1s$ $^2 \mathrm{S} )$ & 9.6003397 & & $^2\Sigma^-$ & $^2\Pi$& \\
  66 & $\mathrm{Y}^{+}(4d7p$ $^3 \mathrm{D}^{\mathrm{o}})+\mathrm{H} ( 1s$ $^2 \mathrm{S} )$ & 9.6049160 & & $^2\Sigma^-$ & $^2\Pi$  & $^2\Delta$ \\
  67 & $\mathrm{Y}^{+}(4d6d$ $^1 \mathrm{G} )+\mathrm{H} ( 1s$ $^2 \mathrm{S} )$ & 9.6244485 &$^2\Sigma^+$& & $^2\Pi$  & $^2\Delta$ \\
  68 & $\mathrm{Y}^{+}(4d7p$ $^3 \mathrm{F}^{\mathrm{o}})+\mathrm{H} ( 1s$ $^2 \mathrm{S} )$ & 9.6285259 &$^2\Sigma^+$& & $^2\Pi$  & $^2\Delta$ \\
  69 & $\mathrm{Y}^{+}(4d7p$ $^1 \mathrm{F}^{\mathrm{o}})+\mathrm{H} ( 1s$ $^2 \mathrm{S} )$ & 9.7570897 &$^2\Sigma^+$& & $^2\Pi$  & $^2\Delta$ \\
  70 & $\mathrm{Y}^{+}(4d7p$ $^1 \mathrm{P}^{\mathrm{o}})+\mathrm{H} ( 1s$ $^2 \mathrm{S} )$ & 9.7689108 &$^2\Sigma^+$& & $^2\Pi$  & $^2\Delta$ \\
  71 & $\mathrm{Y}^{+}(4d6d$ $^1 \mathrm{S} )+\mathrm{H} ( 1s$ $^2 \mathrm{S} )$ & 9.7882053 & $^2\Sigma^+$ & & & \\
  72 & $\mathrm{Y}^{+}(5s7s$ $^3 \mathrm{S} )+\mathrm{H} ( 1s$ $^2 \mathrm{S} )$ & 9.9078994 & $^2\Sigma^+$ & & & \\
  73 & $\mathrm{Y}^{+}(4d5f$ $^1 \mathrm{G}^{\mathrm{o}})+\mathrm{H} ( 1s$ $^2 \mathrm{S} )$ & 9.9467429 & & $^2\Sigma^-$ & $^2\Pi$  & $^2\Delta$ \\
  74 & $\mathrm{Y}^{+}(4d5f$ $^3 \mathrm{F}^{\mathrm{o}})+\mathrm{H} ( 1s$ $^2 \mathrm{S} )$ & 9.9582002 &$^2\Sigma^+$& & $^2\Pi$  & $^2\Delta$ \\
  75 & $\mathrm{Y}^{+}(4d5f$ $^3 \mathrm{H}^{\mathrm{o}})+\mathrm{H} ( 1s$ $^2 \mathrm{S} )$ & 9.9632150 &$^2\Sigma^+$& & $^2\Pi$  & $^2\Delta$ \\
  76 & $\mathrm{Y}^{+}(4d5f$ $^3 \mathrm{G}^{\mathrm{o}} )+\mathrm{H} ( 1s$ $^2 \mathrm{S} )$ & 9.9919338 & & $^2\Sigma^-$ & $^2\Pi$  & $^2\Delta$ \\
  77 & $\mathrm{Y}^{+}(4d5f$ $^1 \mathrm{D}^{\mathrm{o}})+\mathrm{H} ( 1s$ $^2 \mathrm{S} )$ & 10.0053870 & & $^2\Sigma^-$ & $^2\Pi$  & $^2\Delta$\\
  78 & $\mathrm{Y}^{+}(4d5f$ $^3 \mathrm{D}^{\mathrm{o}})+\mathrm{H} ( 1s$ $^2 \mathrm{S} )$ & 10.0207162 & & $^2\Sigma^-$ & $^2\Pi$  & $^2\Delta$ \\
  79 & $\mathrm{Y}^{+}(4d5f$ $^1 \mathrm{F}^{\mathrm{o}})+\mathrm{H} ( 1s$ $^2 \mathrm{S} )$ & 10.0893841 &$^2\Sigma^+$& & $^2\Pi$  & $^2\Delta$ \\
  80 & $\mathrm{Y}^{+}(4d5f$ $^3 \mathrm{P}^{\mathrm{o}} )+\mathrm{H} ( 1s$ $^2 \mathrm{S} )$ & 10.0961782 & $^2\Sigma^+$& & $^2\Pi$  & \\
  81 & $\mathrm{Y}^{+}(4d5f$ $^1 \mathrm{H}^{\mathrm{o}} )+\mathrm{H} ( 1s$ $^2 \mathrm{S} )$ & 10.1147718 &$^2\Sigma^+$& & $^2\Pi$  & $^2\Delta$ \\
  82 & $\mathrm{Y}^{+}(4d5f$ $^1 \mathrm{P}^{\mathrm{o}} )+\mathrm{H} ( 1s$ $^2 \mathrm{S} )$ & 10.1290283 & $^2\Sigma^+$& & $^2\Pi$  & \\
  83 & $\mathrm{Y}^{+}(5s6d$ $^3 \mathrm{D} )+\mathrm{H} ( 1s$ $^2 \mathrm{S} )$ & 10.1909020 &$^2\Sigma^+$& & $^2\Pi$  & $^2\Delta$\\
  84 & $\mathrm{Y}^{+}(4d8s$ $^1 \mathrm{D} )+\mathrm{H} ( 1s$ $^2 \mathrm{S} )$ & 10.2325129 &$^2\Sigma^+$& & $^2\Pi$  & $^2\Delta$ \\
  85 & $\mathrm{Y}^{+}(4d8s$ $^3 \mathrm{D} )+\mathrm{H} ( 1s$ $^2 \mathrm{S} )$ & 10.2464539 &$^2\Sigma^+$& & $^2\Pi$  & $^2\Delta$ \\
  86 & $\mathrm{Y}^{+}(4d7d$ $^1 \mathrm{F} )+\mathrm{H} ( 1s$ $^2 \mathrm{S} )$ & 10.3435355 & & $^2\Sigma^-$ & $^2\Pi$  & $^2\Delta$ \\
  87 & $\mathrm{Y}^{+}(4d7d$ $^3 \mathrm{D} )+\mathrm{H} ( 1s$ $^2 \mathrm{S} )$ & 10.3469480 &$^2\Sigma^+$& & $^2\Pi$  & $^2\Delta$ \\
  88 & $\mathrm{Y}^{+}(4d7d$ $^1 \mathrm{D} )+\mathrm{H} ( 1s$ $^2 \mathrm{S} )$ & 10.3533655 &$^2\Sigma^+$& & $^2\Pi$  & $^2\Delta$ \\
  89 & $\mathrm{Y}^{+}(4d7d$ $^3 \mathrm{G} )+\mathrm{H} ( 1s$ $^2 \mathrm{S} )$ & 10.3624891 &$^2\Sigma^+$& & $^2\Pi$  & $^2\Delta$ \\
  90 & $\mathrm{Y}^{+}(4d7d$ $^3 \mathrm{S} )+\mathrm{H} ( 1s$ $^2 \mathrm{S} )$ & 10.3903980 & $^2\Sigma^+$ & & & \\
  91 & $\mathrm{Y}^{+}(4d7d$ $^3 \mathrm{F} )+\mathrm{H} ( 1s$ $^2 \mathrm{S} )$ & 10.4122324 & & $^2\Sigma^-$ & $^2\Pi$  & $^2\Delta$ \\
  92 & $\mathrm{Y}^{+}(4d7d$ $^3 \mathrm{P} )+\mathrm{H} ( 1s$ $^2 \mathrm{S} )$ & 10.4487930 & & $^2\Sigma^-$ & $^2\Pi$ & \\
  93 & $\mathrm{Y}^{+}(4d7d$ $^1 \mathrm{G} )+\mathrm{H} ( 1s$ $^2 \mathrm{S} )$ & 10.5110796 &$^2\Sigma^+$& & $^2\Pi$  & $^2\Delta$ \\
  94 & $\mathrm{Y}^{+}(5s6d$ $^1 \mathrm{D} )+\mathrm{H} ( 1s$ $^2 \mathrm{S} )$ & 10.5337170 &$^2\Sigma^+$& & $^2\Pi$  & $^2\Delta$ \\
  95 & $\mathrm{Y}^{+}(4d6f$ $^1 \mathrm{G}^{\circ} )+\mathrm{H} ( 1s$ $^2 \mathrm{S} )$ & 10.6503070 & & $^2\Sigma^-$ & $^2\Pi$  & $^2\Delta$ \\
  96 & $\mathrm{Y}^{+}(4d6f$ $^3 \mathrm{F}^{\circ} )+\mathrm{H} ( 1s$ $^2 \mathrm{S} )$ & 10.6562010 &$^2\Sigma^+$& & $^2\Pi$  & $^2\Delta$ \\
  97 & $\mathrm{Y}^{+}(4d6f$ $^3 \mathrm{H}^{\mathrm{o}} )+\mathrm{H} ( 1s$ $^2 \mathrm{S} )$ & 10.6597960 &$^2\Sigma^+$& & $^2\Pi$  & $^2\Delta$ \\
  98 & $\mathrm{Y}^{+}(4d6f$ $^3 \mathrm{G}^{\mathrm{o}})+\mathrm{H} ( 1s$ $^2 \mathrm{S} )$ & 10.6752180 & & $^2\Sigma^-$ & $^2\Pi$  & $^2\Delta$\\
  99 & $\mathrm{Y}^{+}(4d6f$ $^1 \mathrm{D}^{\mathrm{o}})+\mathrm{H} ( 1s$ $^2 \mathrm{S} )$ & 10.6862630 & & $^2\Sigma^-$ & $^2\Pi$  & $^2\Delta$ \\
  100 & $\mathrm{Y}^{+}(4d6f$ $^3 \mathrm{D}^{\mathrm{o}})+\mathrm{H} ( 1s$ $^2 \mathrm{S} )$ & 10.6951870 & & $^2\Sigma^-$ & $^2\Pi$  & $^2\Delta$ \\
  101 & $\mathrm{Y}^{+}(4d6f$ $^1 \mathrm{F}^{\mathrm{o}})+\mathrm{H} ( 1s$ $^2 \mathrm{S} )$ & 10.7647110 &$^2\Sigma^+$& & $^2\Pi$  & $^2\Delta$ \\
  102 & $\mathrm{Y}^{+}(4d6f$ $^3 \mathrm{P}^{\mathrm{o}} )+\mathrm{H} ( 1s$ $^2 \mathrm{S} )$ & 10.7741000 & $^2\Sigma^+$& & $^2\Pi$ & \\
  103 & $\mathrm{Y}^{+}(4d6f$ $^1 \mathrm{H}^{\mathrm{o}})+\mathrm{H} ( 1s$ $^2 \mathrm{S} )$ & 10.7853652 &$^2\Sigma^+$& & $^2\Pi$  & $^2\Delta$ \\
  104 & $\mathrm{Y}^{+}(4d6f$ $^1 \mathrm{P}^{\mathrm{o}})+\mathrm{H} ( 1s$ $^2 \mathrm{S} )$ & 10.7931520 & $^2\Sigma^+$& & $^2\Pi$  & \\
  105 & $\mathrm{Y}^{+}(5s5f$ $^3 \mathrm{F}^{\mathrm{o}} )+\mathrm{H} ( 1s$ $^2 \mathrm{S} )$ & 10.8137400 &$^2\Sigma^+$& & $^2\Pi$  & $^2\Delta$ \\
  106 & $\mathrm{Y}^{+}(4d9s$ $^3 \mathrm{D} )+\mathrm{H} ( 1s$ $^2 \mathrm{S} )$ & 10.8989350 &$^2\Sigma^+$& & $^2\Pi$  & $^2\Delta$ \\
  107 & $\mathrm{Y}^{+}(4d9s$ $^1 \mathrm{D} )+\mathrm{H} ( 1s$ $^2 \mathrm{S} )$ & 10.9046660 &$^2\Sigma^+$& & $^2\Pi$  & $^2\Delta$ \\
  108 & $\mathrm{Y}^{+}(4d8d$ $^3 \mathrm{G} )+\mathrm{H} ( 1s$ $^2 \mathrm{S} )$ & 10.9133660 &$^2\Sigma^+$& & $^2\Pi$  & $^2\Delta$ \\
  109 & $\mathrm{Y}^{+}(4d8d$ $^3 \mathrm{F} )+\mathrm{H} ( 1s$ $^2 \mathrm{S} )$ & 11.0046190 & & $^2\Sigma^-$ & $^2\Pi$  & $^2\Delta$ \\
  110 & $\mathrm{Y}^{+}(5s8s$ $^3 \mathrm{S} )+\mathrm{H} ( 1s$ $^2 \mathrm{S} )$ & 11.0690660 & $^2\Sigma^+$ & & & \\
  111 & $\mathrm{Y}^{+}(4d7f$ $^3 \mathrm{G}^{\mathrm{o}})+\mathrm{H} ( 1s$ $^2 \mathrm{S} )$ & 11.0772100 & & $^2\Sigma^-$ & $^2\Pi$  & $^2\Delta$ \\
  112 & $\mathrm{Y}^{+}(4d7f$ $^3 \mathrm{H}^{\mathrm{o}})+\mathrm{H} ( 1s$ $^2\mathrm{S} )$ & 11.0890600 &$^2\Sigma^+$& & $^2\Pi$  & $^2\Delta$ \\
  113 & $\mathrm{Y}^{+}(4d7f$ $^3 \mathrm{D}^{\mathrm{o}})+\mathrm{H} ( 1s$ $^2\mathrm{S} )$ & 11.0937500 & & $^2\Sigma^-$ & $^2\Pi$  & $^2\Delta$ \\
  114 & $\mathrm{Y}^{+}(4d7f$ $^3 \mathrm{F}^{\mathrm{o}})+\mathrm{H} ( 1s$ $^2\mathrm{S} )$ & 11.1715800 &$^2\Sigma^+$& & $^2\Pi$  & $^2\Delta$ \\
  115 & $\mathrm{Y}^{+}(4d7f$ $^3 \mathrm{P}^{\mathrm{o}})+\mathrm{H} ( 1s$ $^2\mathrm{S} )$ & 11.1834000 & $^2\Sigma^+$& & $^2\Pi$  & \\
  116 & $\mathrm{Y}^{+}(4d7f$ $^1 \mathrm{H}^{\mathrm{o}})+\mathrm{H} ( 1s$ $^2\mathrm{S} )$ & 11.1905200 &$^2\Sigma^+$& & $^2\Pi$  & $^2\Delta$ \\
  117 & $\mathrm{Y}^{2+}(4p^64d$ $^2\mathrm{D} )+\mathrm{H}^{-} (1s^2$ $^1\mathrm{S} )$ & 11.4696000 &$^2\Sigma^+$& & $^2\Pi$  & $^2\Delta$ \\
  \hline
\end{tabular} 
\end{table} 

%\end{supertabular}
%\end{center}
\subsection{Y$^+$+H and Y$^{2+}$+H$^-$ collisions}

\begin{figure*}
	\includegraphics[width=17.3cm]{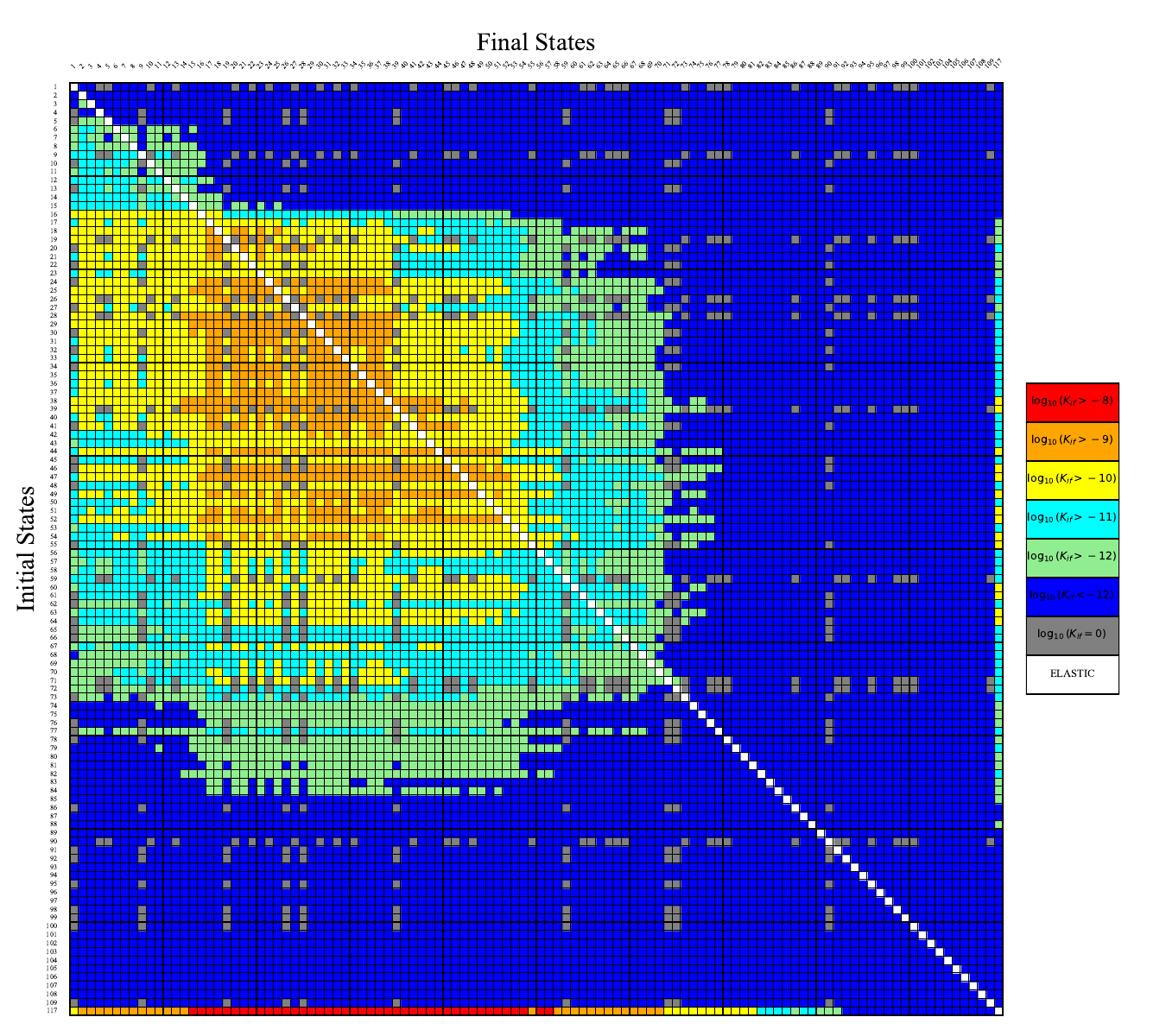}
    \caption{Rate coefficients (in cm$^3s^{-1}$) for the partial excitation, de-excitation, mutual neutralization, and ion-pair formation processes in Y$^{2+}(4p^64d$ $^2\mathrm{D})$+H$^-$ collisions at temperature of T= 6000K. Initial and final state (or scattering channels) labels are presented in Table \ref{table3}.}
    \label{fig4}
\end{figure*}

The ground state Y$^{2+}(4p^64d$ $^2\mathrm{D})$+H$^-$ comprises three molecular symmetries: $^2\Sigma^+$, $^2\Pi$, and $^2\Delta$. In our calculations, we only consider covalent scattering channels that possess these molecular symmetries, which include 116 covalent states of Y$^+[(4d\mathrm{n}l)/(5s\mathrm{n}l)/(5p\mathrm{n}l)$ $^{1,3}L)$+H. The asymptotic atomic states and the corresponding energies of the relevant molecular states are given in Table \ref{table3}. In the same process, the initial and final states may have different molecular symmetries. Therefore, for the calculation of the rate coefficients for inelastic processes, we need to calculate the rate coefficient for each molecular symmetry separately, and then sum these results to obtain the total rate coefficient for the entire process. Using the simplified model, we calculate the rate coefficients for Y$^+$+H collision system for the temperatures between 1000 and 10000K, demonstrating the general trends of the rate coefficients for Y$^+$-H collisions. Same as dealing with the collisions between neutral atoms and hydrogen atoms, only single-electron transitions are considered in this model. 

    Figure \ref{fig4} illustrates the rate coefficients for excitation, de-excitation, ion-pair formation, and mutual neutralization processes in Y$^+[(4d\mathrm{n}l)/(5s\mathrm{n}l)/(5p\mathrm{n}l)$ $^{1,3}L)$+H and Y$^{2+}(4p^64d$ $^2\mathrm{D})$+H$^-$ collisions at T=6000K. Complete data of rate coefficient in the temperatures range between 1000 and 10000K can be found in the supplementary material. In this figure we only display the rate coefficients for the transitions which involving covalent states from $j$=1 to $j$=102 and the ionic states Y$^{2+}(4p^64d$ $^2\mathrm{D})$+H$^-$, due to a fact that the values of rate coefficient for the processes involving covalent states higher than $j$=102 are significantly small. The largest rate coefficients are found in the mutual neutralization processes of Y$^{2+}(4p^64d$ $^2\mathrm{D})$+H$^-$, and the rate coefficients for the final covalent scattering channels denoting as $j$=15-57 (with the exception of $j$=55) exceed  $10^{-8}$cm$^3$s$^{-1}$ (refer to Table  \ref{table3}). Other processes involving ion-pair formation, excitation, and de-excitation, with the values of rate coefficients lying between $10^{-12}-10^{-8}$cm$^3$s$^{-1}$, are denoted by green, cyan, yellow, and orange squares in the same figure. 
The intermediate-to-high values of rate coefficients for the excitation and de-excitation processes are primarily limited to between the initial and final channels marked as $j$=14-54,  and the maximum is the order of $10^{-9}$cm$^3$s$^{-1}$, which is an order of magnitude smaller than the highest rate coefficient of the mutual neutralization process. The results for the elastic processes are represented by white squares on the diagonal in the same figure. Transitions between molecular states with different symmetries (symbolized by grey squares in the same figure) are forbidden, so the values of the rate coefficients are zero.

\begin{figure}
	\centering 
	\includegraphics[width=8.5cm]{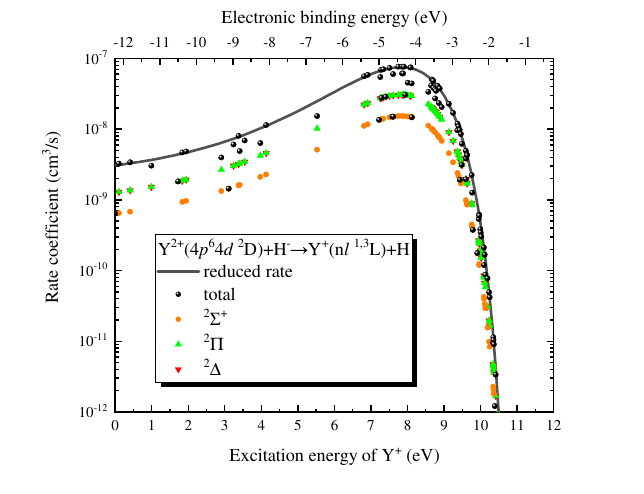}
    \caption{
    Rate coefficients for mutual neutralization processes in Y$^{2+}(4p^64d$ $^2$D)+H$^-$  collisions at T=6000K as a function of the excitation energy of Y$^+$ ions. 
    }
    \label{fig5}
\end{figure}

\begin{figure}
	\centering 
	\includegraphics[width=8.5cm]{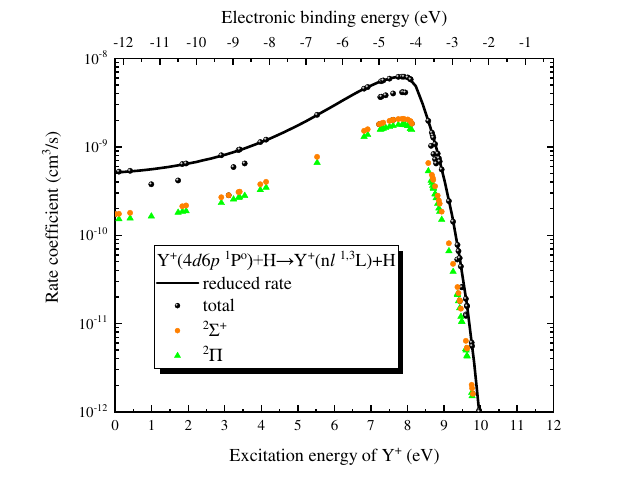}
    \caption{
    Rate coefficients for (de-)excitation processes in Y$^{+}(4d6p$ $^1$P$^o$)+H  collisions at T=6000K as a function of the excitation energy of Y$^+$ ions.
   }
    \label{fig6}
\end{figure}

The mutual neutralization process in Y$^{2+}(4p^64d$ $^2$D)+H$^-$ collisions manifests the largest rate coefficients, as shown in Fig. \ref{fig5}. This figure illustrates the variation of the total rate coefficients for the mutual neutralization process with the excitation energy of Y$^+$ ions at a temperature of T=6000K. The ground ionic state Y$^{2+}(4p^64d$ $^2$D)+H$^-$ as the initial channel has three molecular symmetries ($^2\Sigma^+$, $^2\Pi$ and $^2\Delta$). In this figure, solid lines denote the reduced rate coefficients for the process of singly charged ion (Y$^+$) colliding with hydrogen atoms, which calculated by the simplified model. The figure also displays rate coefficients for different symmetries of initial molecular states, the summation of which yields the total rate coefficients for the corresponding collision process. Furthermore, the figure reveals that the position of optimal window in the collisions between singly charged ions Y$^+$ and hydrogen atoms is proximal to the electron binding energy equivalent to -4.4eV, which corresponding to the excitation energy of Y+ ions are around 7.83eV. The rate coefficient for the final scattering channel which label ed as $j$=31 (Y$^+(4d5d$ e$^1$G)+H) demonstrates a peak of $7.67\times 10^{-8}\mathrm{cm}^3\mathrm{s}^{-1}$ within the mutual neutralization process.

Figure \ref{fig6} shows the variation of total rate coefficients for the excitation and de-excitation processes with the excitation energy of Y$^+$ ions at T=6000K in Y$^{+}(4d6p$ $^1$P$^o$)+H collisions. This corresponds to the process which the covalent state Y$^{+}(4d6p$ $^1$P$^o$)+H is the initial channel, and the molecular ion [YH]$^+$ has two symmetries: $^2\Sigma^+$ and $^2\Pi$. Similar to Fig. \ref{fig5}, this figure also exhibits rate coefficients under different symmetries of the initial molecular ion states. Total rate coefficients can be obtained by the summation of the results from different molecular symmetries. The figure identifies that the position of optimal window also approaches the electron binding energy of -4.4eV. The rate coefficient for the de-excitation process Y$^+(4d6p$ $^1$P$^o$)+H$\rightarrow$Y$^+(4d5d$ e$^1$G)+H ($j$=38$\rightarrow$$j$=31) shows a peak of $6.22\times 10^{-9}\mathrm{cm}^3\mathrm{s}^{-1}$.

\section{Conclusions}  \label{S4}
   The low-energy inelastic collision processes between yttrium atoms (ions) and hydrogen atoms have been investigated using the simplified model for the first time. The rate coefficients for mutual neutralization, ion-pair formation, excitation, and de-excitation processes of the above collision systems have been provided in the temperature range of 1000 to 10000K. For the calculations of Y-H collision system, we consider 3 ionic scattering channels and 73 covalent scattering channels, including 4074 partial inelastic reaction channels. For Y$^+$-H system, 1 ionic scattering channels and 116 covalent scattering channels are included in the calculations, 13572 partial inelastic reaction channels are considered. The computations for different molecular symmetries and spins are treated separately. The total rate coefficients can be obtained by summing the results from different molecular symmetries and spins. It is found that that the rate coefficients from the mutual neutralization processes exhibit the largest value in the considered temperature range.

In the inelastic processes involving the ionic state Y$^+(5s^2$ a$^1$S) which has the molecular symmetry of $^1\Sigma^+$, there are 1332 partial inelastic processes. The rate coefficients for the neutralization process Y$^+(5s^2$ a$^1$S)+H$^-$$\rightarrow$Y$(5s^2(2^\mathrm{D})5d$ e$^2$D)+H shows the largest value of $6.91\times 10^{-8}\mathrm{cm}^3\mathrm{s}^{-1}$. 
For the collision processes involving the initial ionic state Y$^+(4d5s$ a$^3$D), which includes the molecular symmetries of $^3\Sigma^+$, $^3\Pi$, and $^3\Delta$, there are 1482 partial inelastic processes are considered in the present calculations. The rate coefficients for the neutralization process of Y$^+(4d5s$ a$^3$D)+H$^-$$\rightarrow$Y$(5s^2(2^\mathrm{D})5d$ e$^2$D)+H has the largest value of $6.77\times 10^{-8}\mathrm{cm}^3\mathrm{s}^{-1}$.
The inelastic processes related to the ionic state Y$^+(4d^2$ a$^3$F), including four molecular symmetries of $^3\Sigma^-$, $^3\Pi$, $^3\Delta$, and $^3\Phi$, involve 1260 partial inelastic processes. The process of Y$^+(4d^2$ a$^3$F)+H$^-$$\rightarrow$Y$(4d^2(3^\mathrm{F})6s$ $^4$F)+H manifests the highest rate coefficient with a value of $6.95\times 10^{-8}\mathrm{cm}^3\mathrm{s}^{-1}$.

 In the inelastic processes involving the ionic state Y$^{2+}(4p^64d$ $^2$D), we consider three molecular symmetries: $^2\Sigma^+$, $^2\Pi$, and $^2\Delta$. The neutralization process Y$^{2+}(4p^64d$ $^2$D)+H$^-$ $\rightarrow$Y$^+(4d5d$ e$^1$G)+H exhibits the greatest rate value of $6.95\times 10^{-8}\mathrm{cm}^3\mathrm{s}^{-1}$. The de-excitation process Y$^+(4d6p$ $^1$P$^o$)+H$\rightarrow$Y$^+(4d5d$ e$^1$G)+H demonstrates a peak rate value of $6.22\times 10^{-9}\mathrm{cm}^3\mathrm{s}^{-1}$.
 It is found that near the optimal window (-2eV and -4.4eV), excitations, de-excitations, and ion-pair formation processes also have intermediate values of the rate coefficients ($10^{-12}$-$10^{-8}\mathrm{cm}^3\mathrm{s}^{-1}$), which are an order of magnitude smaller than those of the mutual neutralization processes. It is anticipated that these intermediate-to-large rate coefficients will be useful for astrophysical applications. The simplified model provides high accurate and reliable large-value rate coefficients, which is vital for non-LTE modeling. 
 
\section*{Acknowledgements}

This work was supported by the National Key Research and Development Program of China (Grants No. 2022YFA1602500) and National Natural Science Foundation of China (Grants Nos, 11934004, 12274040, 11988101) and the Beijing Natural Science Foundation (grant no. 2192049). S.A.Y. and A.K.B. gratefully acknowledges support from the Russian Science Foundation (the Russian Federation), Project No. 22-23-01181. 

%%%%%%%%%%%%%%%%%%%%%%%%%%%%%%%%%%%%%%%%%%%%%%%%%%
\section*{Data Availability}
The data underlying this article are available as online supplementary material to this paper.

%\bibliographystyle{unsrtnat}
%\bibliography{yh}  %%% Uncomment this line and comment out the ``thebibliography'' section below to use the external .bib file (using bibtex) .

\section*{SUPPORTING INFORMATION}
Supplementary data are available at MNRAS online, supplementary YH.zip.
Rate coefficients in cm$^3$s$^{-1}$ for inelastic Y+H (include Y$^+$+H$^-$) and Y$^{+}$+H (include Y$^{2+}$+H$^-$)  collisions at temperatures from T=1000K to T=10000K.

\end{document}